\documentclass[preprint2]{aastex}

\usepackage{color}

\slugcomment{Accepted for publication in ApJ -- 2011.06.01}

\shorttitle{A 1.2\,mm Map of the Lockman Hole North}
\shortauthors{Lindner et al.}

\begin{document}

\title{A Deep 1.2\,mm Map of the Lockman Hole North Field}

\author{R.~R. Lindner\altaffilmark{1}, 
A.~J. Baker\altaffilmark{1}, 
A. Omont\altaffilmark{2}, 
A. Beelen\altaffilmark{3}, 
F.~N. Owen\altaffilmark{4}, 
F. Bertoldi\altaffilmark{5}, 
H. Dole\altaffilmark{3,6}, \\
N. Fiolet\altaffilmark{2,3}, 
A.~I. Harris\altaffilmark{7}, 
R.~J. Ivison\altaffilmark{8}, 
C.~J. Lonsdale\altaffilmark{9}, 
D. Lutz\altaffilmark{10}, \& 
M. Polletta\altaffilmark{11}}

\altaffiltext{1}{Department of Physics and Astronomy, Rutgers, the State 
University of New Jersey, 136 Frelinghuysen Road, Piscataway, NJ 08854-8019;
\texttt{\{lindner,ajbaker\}@physics.rutgers.edu}}
\altaffiltext{2}{Institut d'Astrophysique de Paris, Universit\'e Pierre Marie 
Curie and CNRS, 98 bis boulevard Arago, 75014 Paris, France; 
\texttt{\{omont,fiolet\}@iap.fr}}
\altaffiltext{3}{Institut d'Astrophysique Spatiale, Universit\'e Paris Sud 11 and 
CNRS, Orsay, France; 
\texttt{\{alexandre.beelen,herve.dole\}@ias.u-psud.fr}}
\altaffiltext{4}{National Radio Astronomy Observatory, PO Box 0, Socorro, NM 
87801; \texttt{fowen@nrao.edu}}
\altaffiltext{5}{Argelander Institut f\"ur Astronomie, Auf dem H\"ugel 71, 
53121 Bonn, Germany; \texttt{bertoldi@astro.uni-bonn.de}}
\altaffiltext{6}{Institut Universitaire de France}
\altaffiltext{7}{Department of Astronomy, University of Maryland, College Park,
MD 20742-2421; \texttt{harris@astro.umd.edu}}
\altaffiltext{8}{UK Astronomy Technology Centre, Royal Observatory, Blackford 
Hill, Edinburgh EH9 3HJ, UK; \texttt{rji@roe.ac.uk}}
\altaffiltext{9}{North American ALMA Science Center, National Radio Astronomy 
Observatory, 520 Edgemont Road, Charlottesville, VA 22903; 
\texttt{clonsdal@nrao.edu}}
\altaffiltext{10}{Max-Planck Institut f\"ur extraterrestrische Physik, Postfach 
1312, 85741 Garching, Germany; \texttt{lutz@mpe.mpg.de}}
\altaffiltext{11}{INAF-IASF Milano, via E. Bassini 15, 20133 Milan, Italy; 
\texttt{polletta@lambrate.inaf.it}}

\begin{abstract} 
We present deep 1.2\,mm continuum mapping of a 566 ${\rm arcmin^{2}}$ 
area within the Lockman Hole North field, previously a target of the {\it 
Spitzer} Wide-area Infrared Extragalactic (SWIRE) survey and extremely deep 
20\,cm mapping with the Very Large Array, which we have obtained using the 
Max-Planck millimeter bolometer (MAMBO) array on the IRAM 30\,m telescope.  
After filtering, our full map has an RMS sensitivity 
ranging from 0.45 to $1.5\,\rm mJy\,beam^{-1}$, with an average of 
$0.75\,\rm mJy\,beam^{-1}$.  Using the pixel 
flux distribution in a map made from our best data, we determine the 
shape, normalization, and approximate flux density cutoff for 1.2\,mm 
number counts well below our nominal sensitivity and confusion limits.  
After validating our full dataset through comparison with this map, 
we successfully detect 41 1.2\,mm sources with ${\rm S/N} > 4.0$ and 
$S_{\rm 1.2\,mm} \simeq 2-5\,$mJy.  We use the most significant of these 
detections to directly determine the integral number counts down to 
1.8\,mJy, which are consistent with the results of the pixel flux 
distribution analysis.  93\% of our 41 individual detections have 
20\,cm counterparts, 49\% have {\it Spitzer}/MIPS $24\,{\rm \mu m}$ 
counterparts, and one may have a significant {\it Chandra} X-ray 
counterpart.  We resolve $\simeq 3\%$ of the cosmic infrared background 
(CIB) at 1.2\,mm into significant detections, and directly estimate a 
0.05\,mJy faint-end cutoff for the counts that is consistent with the 
full intensity of the 1.2\,mm CIB.  The median redshift of our 17 
detections with spectroscopic or robust photometric redshifts is 
$z_{\rm median}=2.3$, and rises to  $z_{\rm median}=2.9$ when we include 
redshifts estimated from the radio/far-infrared spectral index.  By 
using a nearest neighbor and angular correlation function analysis, we find 
evidence that our $\rm S/N>4.0$ detections are clustered at the 95\% 
confidence level.  
\end{abstract}


\section{Introduction} \label{s-intro}

Over a decade ago, measurement of the cosmic infrared background 
\citep[CIB:][]{puget96,fixsen98} revealed that approximately half 
of all of the light in the universe emerges at far-infrared wavelengths due to  
reprocessing by dust \citep[e.g.,][]{dole06}.  With the advent of the Submillimeter 
Common-User Bolometer Array \citep[SCUBA;][]{holl99}, it became clear 
around the same time that not all of this reprocessed emission 
originated in galaxy populations that could be easily detected with 
optical telescopes.  Instead, SCUBA surveys at $850\,\rm\mu m$ revealed 
the existence of a population of bright submillimeter galaxies 
\citep[SMGs;][]{smai97,barg98,hugh98} with faint or undetectable 
optical counterparts.  Optical spectroscopy of the least obscured members 
of the population \citep{ivis98,ivis00}, followed up by detections of CO 
emission \citep{fray98,fray99}, confirmed that SMGs were indeed a 
high-\textit{z} population.  Their faint X-ray counterparts 
\citep{alex03,alex05} as well as mid-infrared spectroscopy
\citep{vali07,mene07,mene09,pope08} indicated that they were not 
predominantly powered by accretion, but rather by star formation.
These observations suggest that SMGs may play an important role in the cosmic star 
formation history.  However, the details of their star formation remain 
uncertain.  Different schools of thought exist about whether SMGs are 
major mergers \citep[e.g.,][]{cons03,nara09,nara10} or host bursts 
triggered by the inflow 
of intergalactic gas along filaments \citep{dave10}.  Likewise, galaxy 
evolution models that consider SMGs have disagreed on whether they 
are \citep{baug05} or are not \citep{hayw11} forming stars with an 
unusually top-heavy initial mass function (IMF).

Understanding how SMGs fit into the overall history of cosmic star 
formation has been impeded by their high obscuration in the optical and 
the coarse angular resolution of the (sub)millimeter bolometer arrays used to 
detect them.  An important advance came with the use of deep, high-resolution 
radio continuum maps with the Very Large Array (VLA) to localize SMGs on 
the basis of the far-IR/radio correlation.  Keck followup of radio-preselected 
SMG samples \citep{ivis02,chap03,chap05} broke the logjam for 
determining 
SMGs' redshifts, allowing a dramatic expansion of SMG samples with CO-confirmed 
spectroscopic redshifts and spatially resolved maps \citep{neri03, grev05,tacc06}.  
Although these 
developments have been important and exciting, there are a number of 
caveats on the current state of our knowledge.  First, not all SMGs have 
counterparts in VLA maps of typical depths, and not all that have counterparts 
yield optical spectroscopic redshifts; this has led to uncertainties in 
the population's overall redshift distribution, especially at the high-$z$ 
end.  Second, we have come to realize that the $850\,\rm\mu m$ waveband 
at which the influential early SCUBA work was done does not give us a 
complete picture of all dusty galaxy populations at high redshift: 
observations at shorter/longer wavelengths preferentially detect 
populations with higher/lower dust temperatures and/or lower/higher 
redshifts \citep[e.g.,][]{chap04,vali07,magn10,chap10,magd10}.  Finally, with 
the exception of highly lensed systems, our direct knowledge is restricted to 
bright individual detections for which limited sensitivity and angular 
resolution (vs. confusion) are not fatal obstacles.  Understanding the 
lower-$L_{\rm IR}$ galaxy populations that produce the bulk of the CIB 
is challenging, and efforts to determine their properties are not always 
consistent with the CIB's normalization \citep[e.g.,][]{scot10}.

To fill in some of the gaps in our knowledge, we need to obtain deep 
mapping at multiple (sub)millimeter wavelengths, at high angular 
resolution, over a large area that has good multiwavelength coverage 
(and especially, very deep radio continuum data).  This combination 
of properties can in principle allow us to (a) optimize the identification 
of counterparts at other wavelengths, and therefore the determination of 
redshifts and the assessment of SMGs' detailed evolutionary states; (b) 
defeat cosmic variance and optimize sensitivity to clustering and 
large-scale structure, a key first step in connecting the properties of 
SMGs to the properties of the dark matter halos that host them; and (c) 
strongly constrain the parameters of SMG number counts down to faint 
flux densities, so that we can accurately compare the census of obscured 
star formation to the constraint of the CIB.  

In this paper, we report $1.2\,\rm mm$ observations at $11''$ resolution 
of a unique deep field that satisfies many of these needs.
Our map is both larger and more sensitive than previous deep maps 
made at $1.2\,\rm mm$ \citep[e.g.,][]{grev04,bert07,grev08}, and
compared to recent work done with other instruments, displays a 
competitive combination of attributes.  
Surveys made at $1.1\,\rm mm$ using ASTE/AzTEC have mapped 
equal or wider fields to a greater depth, but with much lower 
spatial resolution \citep[e.g.,][]{scot10,hats10}.  Maps using JCMT/AzTEC 
and APEX/LABOCA generally achieve wider fields or greater depths, but not both,
and also have coarser spatial resolution 
\citep[e.g.,][]{scot08,perer08,weiss09,aust10}.  
The powerful combination
of resolution, depth, and extent of our MAMBO map, together with 
the rich supplementary data available for our target
field, make it a powerful tool for studying the properties of 
SMGs.

Our map lies within the Lockman Hole North (LHN) field, centered at 10:46:00 and 
+59:01:00 (J2000), which was one of the targets of the {\it Spitzer} 
Wide-Area Infrared Extragalactic (SWIRE) survey \citep{lons03}.  In addition 
to SWIRE coverage in all four IRAC and all three MIPS bands, the LHN has been the 
subject of extremely deep 20\,cm continuum mapping with the Very Large Array 
(VLA) by \citet{owen08}, who produced an ABCD configuration synthesis image 
with a central $1\sigma$ RMS sensitivity of $2.7\,{\rm \mu Jy}$.  These data 
have been further supplemented by 90\,cm VLA mapping \citep{owen09a}, 50\,cm
GMRT mapping \citep{fiol09,owen11a}, deep \textit{Spitzer} $24\,\rm\mu m$ 
imaging \citep{owen11b}, optical spectroscopy with WIYN 
\citep{owen09a}, X-ray imaging from the \textit{Chandra}/SWIRE survey 
\citep{poll06,wilk09}, and determination of photometric redshifts from multicolor 
optical and near-IR imaging \citep{stra10}.  The LHN is also
one of the targets of the \textit{Herschel} Multi-tiered Extragalactic Survey 
\citep[HerMES;][]{oliv10}.

\section{Observations} \label{s-obs}

We used the 117-element Max-Planck millimeter bolometer (MAMBO) array 
\citep{krey98} at the IRAM 30\,m telescope to obtain on-the-fly mapping of the 
LHN at an effective wavelength of 1.2\,mm.  Our observations were 
obtained over the course of five semesters from 2006 through 2010 (Table 
\ref{t-obs}).  Due to telescope control software problems during the first 
two semesters -- an error in computing corrections for atmospheric refraction, 
which undermined the quality of the pointing during the first semester, and 
tracking jitter that undermined map reconstruction during both semesters -- we 
have restricted our initial analysis (e.g., \S \ref{s-pd}) to a ``best'' map 
that includes only the data from our final three semesters of observations.  
We have also constructed a ``full'' map using data from all five semesters, 
whose reliability we can validate based on comparison with the ``best'' map 
(\S \ref{ss-hinge}), and which we therefore use for our analysis of the bright 
source population in the field.  101.3\,hr and 192.5\,hr of data were combined
to produce the ``best'' and ``full'' maps with areas of 
$514\,\rm arcmin^{2}$ and $566\,\rm arcmin^{2}$, and average depths of
$0.90\,\rm mJy\,beam^{-1}$ and $0.75\,\rm mJy\,beam^{-1}$, respectively.

During all five semesters, our MAMBO data were acquired during the weeks 
that IRAM dedicates to pooled observations of multiple bolometer projects.  As 
a result, most of our data were taken with 250\,GHz zenith opacity $\leq 0.3$, 
low sky noise, and essentially no cloud cover.  Observations were limited to 
LST ranges when the LHN had elevation $\geq 40^\circ$ (to minimize opacity 
corrections and pointing anomalies) and $\leq 80^\circ$ (to avoid slewing 
errors and sudden accelerations of the MAMBO array).  We built 
up coverage of 
our field by making many small $\sim 300^{\prime\prime} \times 
320^{\prime\prime}$ (azimuth $\times$ elevation) maps, each of which 
required 41 minutes to complete.  The pointing centers of these small maps 
were arranged in a rectangular grid with $2^\prime$ spacings between map 
centers.  

We planned the observations so that before (and usually after) each 41-minute 
map, the telescope was pointed on a nearby quasar-- typically J1033+609
at a distance of $1.9^{\circ}$ from field center, 
with $S_{1.2\rm\, mm}\sim 0.3\,{\rm Jy}$.  Every 1--2 maps, a skydip was performed to 
measure the zenith opacity, confirm that weather conditions remained good, 
and allow the interpolation of opacity corrections during the maps themselves. 
Standard flux calibrators were observed roughly every four 
hours; these coincided with large slews, to force resets of the telescope's 
inclinometers.  Focus measurements on bright quasars or planets were made 
at the beginning of each observing session as well as after sunrise and 
sunset.  We required all of these calibrations so as to minimize pointing 
errors and anomalous refraction, as is important for the detection of faint 
point sources in a wide-area map.

The IRAM 30\,m uses a chopping secondary mirror to subtract low-frequency sky 
noise from on-the-fly MAMBO maps.  This chopping gives the telescope an 
effective double-beam point spread function (PSF) on the sky, with one 
positive and one negative beam separated in azimuth and symmetric about the 
nominal pointing center.  During shift-and-add (SAA) reconstruction (see \S 
\ref{s-red}), negative-beam data are inverted and combined (for a given sky 
position) with their positive-beam counterparts.  The end result is a 
triple-beam pattern that is a well-defined function of position for any single 
observation: two negative sidelobes bracket a positive beam in azimuth.  The 
SAA algorithm thus conserves the mean flux of the observations, in that the 
negative sidelobes together contain as much integrated flux as the positive 
beam.  When many observations are combined into a single mosaicked image, the 
effective PSF is a superposition of many triple-beam patterns that can vary 
substantially with position.  

Since a given sky position within the LHN usually falls within several of our 
small maps, varying the chop throw and scan direction tends to scatter the 
negative flux into a uniform annulus around the positive central Gaussian, 
reducing its peak intensity and minimizing its deleterious effects on the 
fluxes of nearby pixels.  We therefore (a) used different chop throws for 
alternating columns in our grid of pointing centers, and (b) recorded the 
scan direction of each map in equatorial coordinates, so that observers (to 
the best of their ability) could observe new grid positions at LSTs such 
that scan directions would not match those of (existing) adjacent maps.  In 
the central regions of our final maps (i.e., where we extract sources), peak 
intensities of the negative sidelobes reach only 5$\%$ of the peak positive 
flux thanks to our adoption of these strategies.

\begin{figure} 
\epsscale{1}
\plotone{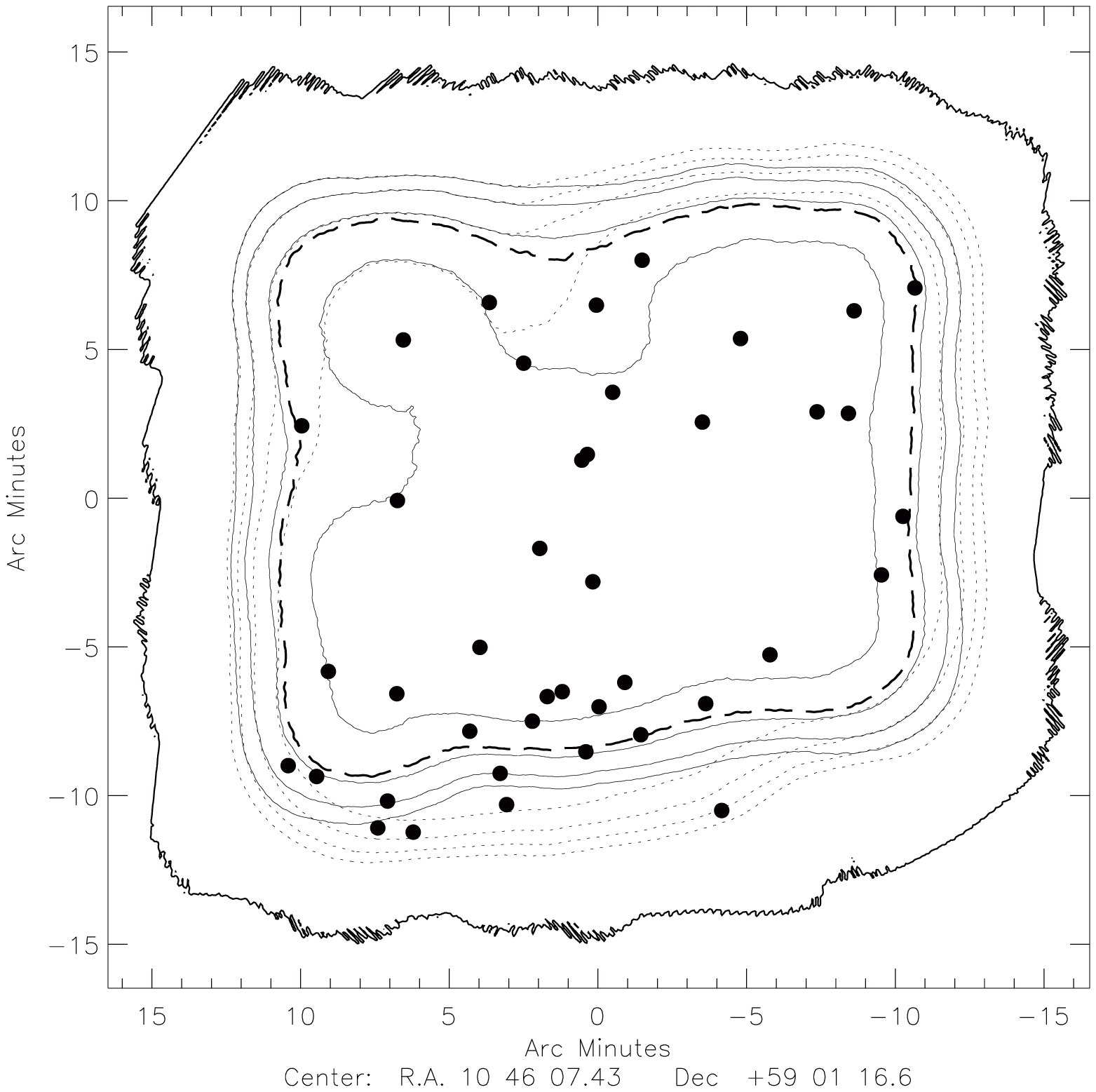}
\caption{Weight map.  Contours denote post-filtering RMS noise levels of 
0.8, 1.0, 1.2, and $\rm 1.4\,mJy\,beam^{-1}$ in the ``best'' data (solid 
contours) and the ``full'' data (dotted contours). Thick dashed contour shows 
the map region used for the \textit{P(D)} analysis of the ``best'' data.  
Circles show the locations of detected sources with ${\rm S/N} \geq 4.0$.  
The thick outer edge shows the extent of the full map.  The effective areas 
comprising the ``best'' and ``full'' maps (RMS noise $<1.5\,\rm mJy\,beam^{-1}$
after filtering) are 514 and $566\,\rm arcmin^{2}$, respectively.}
\label{f-weightmap}
\end{figure}

During our first two semesters of observations, we obtained maps of 88 
grid positions out of 100 in a $10 \times 10$ grid centered on the LHN 
field center stated in \S \ref{s-intro}.  During our last three semesters, 
which contribute to the ``best'' map, we observed 97 positions of 99 in a 
$9 \times 11$ grid, extending $2^\prime$ farther east but $2^\prime$ less far 
south than the original grid, as well as two additional grid locations in the 
southeast corner.  Between the first and second semesters, we swapped which 
sets of pointing centers were observed with which chop throws 
($36^{\prime\prime}$ and $48^{\prime\prime}$ throws vs. $42^{\prime\prime}$ 
and $36^{\prime\prime}$ throws for alternating columns).  Due to the 
differences in spatial coverage and weather conditions during the 
observations, the areas where the ``best'' map and the ``full'' map are 
respectively deepest overlap but do not match perfectly 
(see Figure \ref{f-weightmap}).

\section{Data reduction} \label{s-red}

\subsection{Signal maps} \label{ss-sigmap}

\begin{figure} 
\epsscale{1.}
\plotone{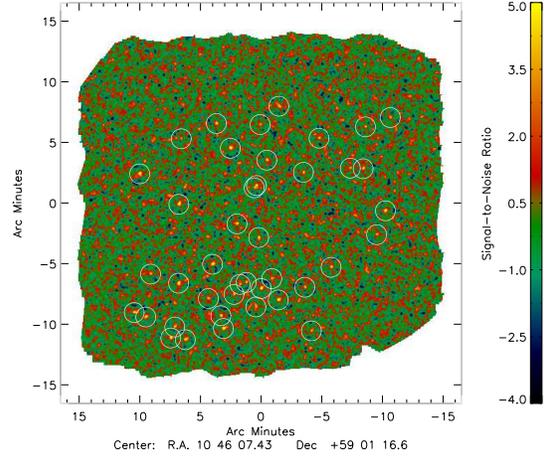}
\caption{The complete, optimally filtered S/N map of the 
``full'' dataset, with white circles showing the locations of our 
41 detections with $\rm S/N > 4.0$.}
\label{f-map}
\end{figure}

The raw bolometer time stream data were reduced using Robert Zylka's 
MOPSIC\footnote{see {\tt http://www.iram.es/IRAMES/mainWiki/CookbookMopsic}}
pipeline, which is distributed in parallel 
with IRAM's GILDAS package. 
MOPSIC is the standard package for reducing deep MAMBO
on-the-fly maps \citep[see e.g.,][]{grev04,voss06,bert07,grev08}.  We now briefly
outline the steps of the MOPSIC reduction pipeline; for further details 
see \cite{grev04}. The 
pipeline removes spikes in the time stream data stronger than $5 \times$ the 
instantaneous bolometer RMS noise.  It also subtracts a third-order 
polynomial baseline in 
time and performs correlated signal filtering on the bolometer time streams to 
identify and remove foreground atmospheric emission that affects many 
bolometers simultaneously.  Each bolometer is correlated with an 
annulus of neighboring
bolometers within a $150''$ radius, and the average signal of the twelve
most highly correlated bolometers is subtracted away.  The filtered time 
streams are 
then binned into 
$3.5^{\prime\prime} \times 3.5^{\prime\prime}$ pixels, and a signal map is 
reconstructed using the SAA algorithm.  The individual signal maps are combined 
into a mosaic image by averaging the map flux density at each pixel weighted 
by the local inverse variance.  Our ``optimally filtered'' signal map 
(Figure \ref{f-map}) was created by applying a PSF-matched filter to the 
final mosaic image (\S \ref{ss-ext}).

In addition to the signal map, the MOPSIC pipeline also produces a weight 
map that is locally proportional to the inverse variance in the signal map.  
By enforcing that the Gaussian distribution of the S/N map pixel distribution 
has a standard deviation of unity, we normalize the weight map so that it can 
be used to find the local RMS noise, $\sigma=1/\sqrt{W}$, in the image (see 
Figure \ref{f-weightmap}).  Using the weight map as a guide to find the local 
RMS noise for a detection is more robust than using the nearby pixels 
themselves, because locally the pixels are affected by the negative residual 
sidelobes of SAA reconstruction, as well as those of other bright nearby 
sources.  

\subsection{Noise maps} \label{ss-noimap}

Because of the telescope's effective triple-beam PSF, each source in the 
field injects negative as well as positive flux into the map.  To generate 
realizations of source-free maps, hereafter referred to as ``noise maps,'' we 
removed the negative and positive flux from undetected as well as bright 
sources using two different techniques.  We go on to use the different 
results for different purposes.

We constructed the first type of noise map with a technique common in MAMBO 
data analysis \citep[see, e.g.,][]{grev04,bert07,grev08}, using the data 
reduction pipeline to scramble the known locations of the bolometers within the
image plane.  During reconstruction of the time stream data, this 
has the effect of smearing the flux from any one source into an area on the 
sky of approximately $200\,{\rm arcmin}^2$, reducing the intensity of the 
source's peak flux contributions by a factor of $\sim 10^3$ and making the 
peak flux contribution from our strongest sources $\sim 200$ times fainter 
than the RMS noise.  Because the telescope's chopping ensures that the mean 
of the map is zero, there is no residual baseline increase as the negative 
flux contributions are identically smoothed.  These ``shuffled noise maps'' 
are simple to construct, but it is cumbersome to produce large numbers of them 
since each requires a full reduction of the data using the MOPSIC pipeline.  
Therefore, we use the shuffled noise maps only to estimate the noise of our 
``full'' data during source extraction (\S \ref{ss-ext}) as well as in the 
Monte Carlo simulation of completeness (\S \ref{ss-comp}).

We needed to develop a different technique for creating noise maps in order 
to quickly generate thousands of independent noise realizations of chopped 
data for our \textit{P(D)} analysis.  For this we subtracted subsets of the 
data that are ``jackknifed'' in the sense that we remove fractions of the 
original data first.  One \textit{full} image of our field is created using 
the data from only \textit{one} bolometer in the array at a time.  All 
bolometers other than the one of interest are masked away after the correlated 
signal filter is applied, so the data still receive the benefit of correlated 
sky noise subtraction.  Two half-sets of these images are then selected at 
random and subtracted from each other to produce one realization of noise.  
This technique is similar to the jackknifing by scan commonly used for AzTEC 
data \citep[see, e.g.,][]{scot08,scot10,perer08,aust10}, in that each 
jackknifed subset uses the scanning information of every available map.  Use
of this information is especially important for our chopped data if we are 
to remove negative flux artifacts from the triple-beam PSF as well as 
positive flux.  These ``jackknifed noise maps'' are more amenable to mass 
production, and are guaranteed to remove all contributions from a source 
however faint, so they are used in our pixel flux distribution (PFD) analysis 
(\S \ref{s-pd}) and in our Monte Carlo simulations to estimate 
numbers of spurious detections (\S \ref{ss-spur}).  

\begin{figure} 
\epsscale{1.}
\plotone{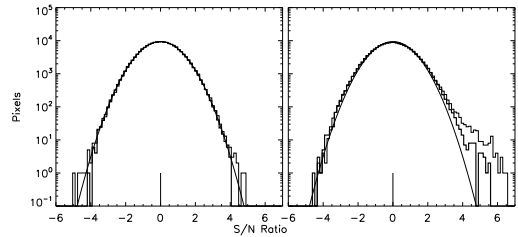}
\caption{Pixel flux distributions of the S/N maps.  Left: Histograms 
are shown for both shuffled (thin) and jackknifed (thick) noise maps.  
  Right: S/N map histogram for the ``best'' 
map (thick) and the ``full'' map (thin).  Over-plotted in both panels is a 
Gaussian function with unit standard deviation and zero mean, normalized 
to the area under the histograms.  All maps have a mean value consistent 
with zero (shown as the vertical line segment), enforced by the chopped 
observing mode of the IRAM 30\,m telescope and SAA reconstruction. 
The ``full'' map has more pixels with high S/N because
it reaches a higher sensitivity.  The histograms were created with maps 
trimmed to a noise level of $1.5\,\rm mJy\,beam^{-1}$.}
\label{f-histo}
\end{figure}

The PFDs for S/N maps created with both jackknifed 
and shuffled noise exhibit random Gaussian noise to high precision (Figure 
\ref{f-histo}), with reduced chi-square for standard normal distribution 
fits of $1.0 \pm 0.2$ and $1.2 \pm 0.2$, respectively.  

\subsection{Simulated maps} \label{ss-simmap}

Our simulated sky maps are constructed by populating noise maps with 
simulated sources.  Careful construction of these maps is important for the 
fluctuation analysis described below (\S \ref{s-pd}), for which our method 
relies entirely on our ability to authentically reproduce the signal from the 
MAMBO array so as to faithfully reproduce the PFD.  Thus, when adding sources 
into a noise map, we need to take into account the position-dependent negative 
sidelobes as well as the position-independent positive flux profile for each 
injected source.

\begin{figure} 
\epsscale{1.}
\plotone{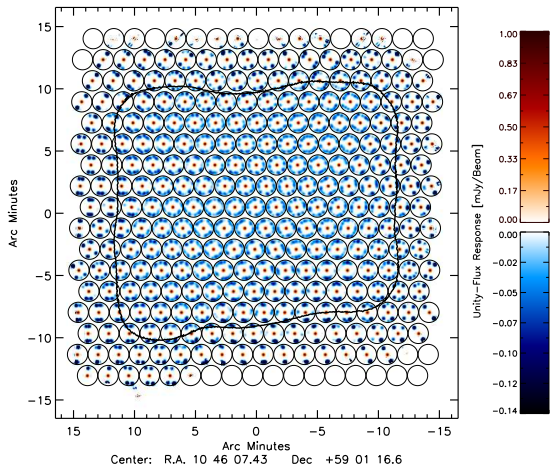}
\caption{The effective PSF (including negative residuals) as a function of 
position across our ``best'' map.  Red/blue represents a positive/negative 
signal response. The thick solid contour encloses the area where
the ``best'' map has RMS noise $<1.5\,\rm mJy\,beam^{-1}$.}
\label{f-negres}
\end{figure}

To handle the varying PSF properly, we take an approach similar to that of 
\cite{grev08} and model the changes in the PSF explicitly as a function of 
position.  We use the MOPSIC pipeline script \texttt{map\_negres.mopsic}, 
which will calculate the expected negative residual pattern on the sky in 
equatorial coordinates for a given set of observations and 
an ideal, gridded, input source model.  As an input we used an array of 
ideal Gaussian point source profiles, 
each with $11^{\prime\prime}$ FWHM, spanning the entire field and spaced as 
closely as possible without having the sidelobes overlap.  This minimum 
spacing is set by our larger chop throw ($42^{\prime\prime}$ for all of our 
``best'' data and the overwhelming majority of our ``full'' dataset).  The 
result is an array showing the full PSF near any location in the map (see 
Figure \ref{f-negres}).  Because the $\sim 84^{\prime\prime}$ spacing is less 
than the $300^{\prime\prime}$ extent of each individual map and the 
$120^{\prime\prime}$ separation between map pointing centers, the PSF 
morphologies change slowly from one to the next.  We thus generate an 
authentic point source response in a simulated map by using the closest 
available PSF relative to the position of a given injected source.

\section{\textit{P(D)} analysis of the pixel flux distribution} \label{s-pd}

\begin{figure} 
\epsscale{1.}
\plotone{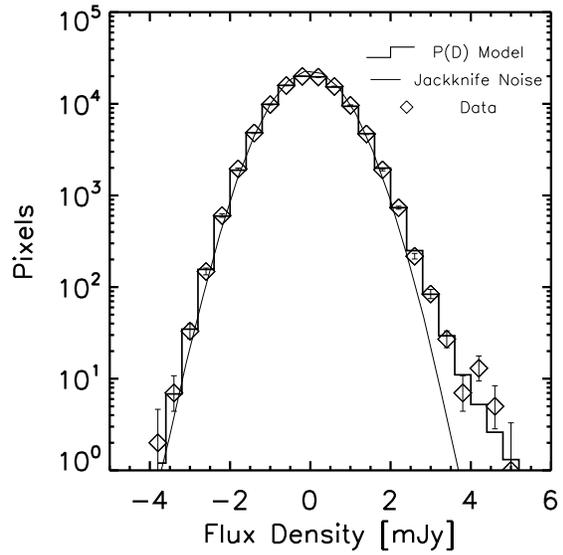}
\caption{Pixel flux distribution showing the agreement between
the data and the average best-fit (Schechter function) model from the \textit{P(D)} analysis.   
Points represent the PFD of the optimally filtered 
``best'' data. The solid histogram shows the mean PFD of 
100 random sky realizations of the best fitting number counts embedded in 
random jackknifed noise maps.  The thin curve 
represents the mean histogram of only the jackknifed noise maps.  The
single larger flux density bin that spanned 4--5\,mJy for the $P(D)$ analysis
(see \S \ref{s-pd}) is shown here broken down into small bins matching 
the rest of the distribution.
}
\label{f-PDspace}
\end{figure}

We constrained the 1.2\,mm number counts below our nominal sensitivity and 
confusion limits by performing a fluctuation analysis, also known as a $P(D)$ 
analysis \citep{cond74}.  The $P(D)$ analysis has the advantage of using 
information from the entire PFD of the map (see Figure \ref{f-PDspace}) to 
constrain the number counts, rather than using only those pixels above the 
source detection threshold (e.g., by counting bright sources).  This 
distinction makes the $P(D)$ analysis robust against the small number 
statistics of counting detections in the map.  Additionally, the nature of the 
Monte Carlo simulation described below allows us to minimize uncertainties in 
flux boosting and completeness, as well as the effects of confusion and source 
blending, because they are built into the simulation through the injection of 
model sources.  These benefits have led Monte Carlo simulation \textit{P(D)} 
analyses \cite[e.g.,][]{malo05,scot10}, as well as Markov Chain Monte Carlo 
Metropolis-Hastings (MCMCMH) \textit{P(D)} analyses 
\cite[e.g.,][]{pata09,vali10,glen10}, to be applied to both chopped and 
unchopped data at many wavelengths.  Our implementation of a Monte Carlo 
simulation \textit{P(D)} analysis, which is best suited to handle 
our position-dependent PSF,  adopts the methods of \cite{scot10}.

The basic approach of our $P(D)$ analysis is to parametrize the differential 
number counts and add a simulated map of sources obeying these number counts 
(along with their position-dependent negative sidelobes; see \S 
\ref{ss-simmap}) to a jackknifed noise map (see \S \ref{ss-noimap}), thereby 
creating a fully simulated MAMBO sky image.  Because of its
simple form, our initial parametrization is a 
single power law with normalization $N_{\rm 4\,mJy}$ and index $\delta$, such
that the differential number counts have the form

\begin{equation}
{\frac {dN}{dS}} = N_{\rm 4\,mJy} \left({\frac{4\,{\rm mJy}}
{S}}\right)^\delta
\end{equation}
We adopt this form from \cite{laur05}, so as to minimize the degeneracy 
between the normalization and slope of the number counts at the flux density 
of our typical significant detection ($\simeq 4\,\rm mJy$).  Next, we 
optimally filter this fully simulated sky image and compare its PFD to that of 
the real data using the likelihood (see below) as a goodness-of-fit statistic.
We then compute the average likelihood of the data for ten iterations of 
these model parameters, choose new parameters, and repeat the process.  
After filling parameter space with likelihood statistics, we identify the 
best-fit parameters as those giving the maximum likelihood.  After the 
location of this peak in parameter space is identified, we return 
and sample this one position $\sim 10^{3}$ times in order to constrain 
the absolute likelihood value enough to discriminate between fits using 
different flux density cut-off values (see below).

The likelihood for each sky realization is calculated as follows.  Assuming 
the PFD's flux bins are uncorrelated, the probability of observing $n_i$ 
pixels in the $i$th flux bin given an expectation value of $\lambda_i$ is 
given by a Poisson distribution: 
\begin{equation}
P(n_i|\lambda_i) = {\frac {\lambda_i^{n_i} e^{-\lambda_i}}{n_i!}}
\end{equation}
Therefore, the natural logarithm of the probability $P[n_i]$ of observing the 
full PFD $\{n_i\}$ for a model PFD $\{\lambda_{i}\}$ (the log-likelihood) 
is given by
\begin{equation}
{\rm ln}\,P[n_i] = \sum_i {\rm ln}\,\Big({\frac {\lambda_i^{n_i} 
e^{-\lambda_i}}{n_i!}}\Big) = \sum_i n_i\,{\rm ln}\,\lambda_i - {\rm 
ln}\,n_i! - \lambda_i
\end{equation}
We limit the comparison to bins in which $n_i \ge 10$ and use 
Stirling's approximation to write the sum as
\begin{equation}
{\rm ln}\,P[n_i] \simeq \sum_i n_i - \lambda_i - n_i\,{\rm ln}\,\Big({\frac 
{n_i}{\lambda_i}}\Big)
\end{equation}
In reality our histogram bins are \textit{not} uncorrelated, since our beam 
solid angle is $\sim10\times$ larger than the area of one pixel; thus,  
this expression will serve simply as a comparative metric for choosing a set of 
best-fitting parameters.  The properly calibrated error bars for this estimate 
can then be found via a Monte Carlo simulation using synthetic images generated
from the best-fit model.  $P[n_i]$ is therefore a function over the 
two-dimensional parameter space of ($N_{\rm 4\,mJy}$,$\delta$), within which 
the best-fitting model parameters are those that minimize $-{\rm ln}\,P[n_i]$.

We apply the \textit{P(D)} analysis to the region in the ``best'' map where 
the local RMS noise $\sigma \le 1.25\,{\rm mJy\,beam^{-1}}$ before filtering 
(see Figure \ref{f-weightmap}).  This threshold was chosen to maximize 
the discriminating power of the simulation.  If the noise 
threshold is very low, the region used for analysis has very high 
sensitivity, but there are fewer pixels available for comparison.  If the 
noise threshold is too high (e.g., we use the full extent of the ``best'' map),
 too many regions with differing local noise properties are included, and 
the signal from the interior of the map is washed out.  
Our choice of threshold represents a compromise between these two limits, 
including as many pixels in the analysis as possible while keeping their 
noise properties as uniform as possible.

Our initial expression of the PFD in terms of 15 bins between $-3.5$ and 
$+4.0\,{\rm mJy\,beam^{-1}}$, chosen so that all flux 
bins had $\ge 10$ pixels, failed to constrain the model parameters with a 
unique maximum likelihood.  Because the brightest (and most 
model-constraining) pixels in the histogram are in the sparsely populated bins 
above $4.0\,{\rm mJy\,beam^{-1}}$, our simulations produced only a 
best-fit {\it arc} in parameter space.  When we increased the sensitivity to 
the brightest pixels by adding an additional bin spanning from $4.0-5.0\,{\rm
 mJy\,beam^{-1}}$, wide enough to include $\ge 10$ pixels, a well-defined 
global maximum appeared along the previously degenerate arc.  

In order to keep the models from diverging at low flux 
densities, we also imposed a faint-end cutoff in flux density, $S_{\rm cut}$, 
which we crudely treated as a  third parameter in the $P(D)$ 
analysis.  By stepping through the values $S_{\rm cut}= 0.3$, 0.2, 0.1, 0.05, 
and 0.01\,mJy, testing each with a full set of fitting parameters, we
found the  maximum likelihood values for the power law to be 
 $41.7\pm 0.4\%$, $47.7\pm 0.2\%$, $52.1\pm 0.2\%$, $53.0\pm 0.3\%$, 
and $49.7\pm 0.5\%$, respectively.  The fits improved 
steadily with decreasing $S_{\rm cut}$ down to 0.05\,mJy but then worsened 
at 0.01\,mJy.  Thus, the overall best fitting parameters for the power law were 
$N_{\rm 4\,mJy}$ = $\rm 19.7^{+4.1}_{-8.8}\,{\rm deg^{-2}\,mJy^{-1}}$, 
$\delta = 3.14^{+0.14}_{-0.18}$, and $S_{\rm cut}=0.05\,\rm mJy$ (we quote 
marginalized $68\%$ double-sided error bars).

Our second number counts parametrization was a \citet{sche76} 
function of the form
\begin{equation}
\frac{dN}{dS} = N_{\rm 4\,mJy}'\,\left(\frac{4\rm\,mJy}{S}\right)^{\delta'} 
{\rm exp}\left(-\frac{S - 4\,{\rm mJy}}{S^{\prime}_{\rm exp}}\right)
\end{equation}
Because the full four-dimensional parameter space of the Schechter function
($N_{\rm 4\,mJy}'$, $\delta'$, $S^{\prime}_{\rm exp}$, $S_{\rm cut}$)
is too large to probe with a blind grid-searching routine, we began by
fixing $S_{\rm cut}^{\prime}$ equal to the solution for the power-law 
model ($0.05\,\rm mJy$). We then alternated the \textit{P(D)} analysis 
between varying the parameters ($N_{\rm 4\,mJy}'$, $S'_{\rm exp}$) and 
($N_{\rm 4\,mJy}'$, $\delta'$) until the solutions converged on the same 
values for all three parameters (convergence was achieved after 
three iterations).  The initial seed guess for $S'_{\rm exp}$ was 
motivated by naively scaling the \textit{P(D)} solution at 1.1\,mm, 
found by \cite{scot10} using AzTEC data in the GOODS-S field 
($S_{\rm exp,\,1.1\, mm}'=1.30\,\rm mJy$), to $1.2\,\rm mm$ 
(see \S \ref{ss-compare counts}).  The results converged to the exponential scaling 
flux density of $S_{\rm exp}^\prime = 1.05\,\rm mJy$, which was then held fixed
while we proceeded to make full searches over the parameters 
($N_{\rm 4\,mJy}'$, $\delta'$) while varying $S_{\rm cut}'$.  

The quality-of-fit for the Schechter function was also greatest 
for a flux density cutoff of $S_{\rm cut}'=0.05\,\rm mJy$.  The 
maximum likelihood values for $S_{\rm cut}'=0.2$, 0.1, 0.05, and 
$0.01\,\rm mJy$ were $23.8\pm0.1\%$,  $26.9\pm0.1\%$,  $28.3\pm0.2\%$, 
and  $26.7\pm0.4\%$, respectively.  The final set of best-fitting 
parameters for the Schechter function were 
$N_{\rm 4\,mJy}' = 14.5^{+7.1}_{-2.7}\,\rm deg^{-2} mJy^{-1}$,  
$\delta' = 1.86^{+0.20}_{-0.23}$, $S_{\rm exp}^\prime = 1.05\,\rm mJy$, and
$S_{\rm cut}^{\prime}=0.05\,\rm mJy$.

\begin{figure} 
\epsscale{1.}
\plotone{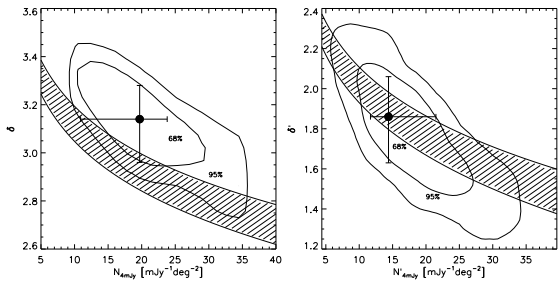}
\caption{Likelihood parameter spaces for 
different parametrizations of the $1.2\,\rm mm$ differential 
number counts.  The points mark the best fitting
parameters and the marginalized $68\%$ double-sided error bars.  The 
contours bound the 68$\%$ and $95\%$ confidence 
regions found through Monte Carlo simulations.  The shaded bands show 
the regions in parameter space that reproduce the observed intensity 
of the $1.2\rm\,mm$ CIB, assuming a lower-limit flux density cutoff 
of $0.05\,\rm mJy$.  \textit{Left}: Power law model, with 
best fitting parameters  
$N_{\rm 4\,mJy} = \rm 19.7^{+4.1}_{-8.8}\,deg^{-2}\,mJy^{-1}$ and 
$\delta = 3.14^{+0.14}_{-0.18}$.
\textit{Right}: Schechter function model, with best fitting parameters
$N_{\rm 4\,mJy}' = 14.5^{+7.1}_{-2.7}\,\rm deg^{-2} mJy^{-1}$ and 
$\delta' = 1.86^{+0.20}_{-0.23}$.}
\label{f-PDerror}
\end{figure}

Figure \ref{f-PDerror} shows full 68\% and 95\% confidence regions around 
the parameters of maximum likelihood for both the power law and Schechter
function parametrizations.  The uncertainty contours were generated via 
Monte Carlo sampling \citep[used, e.g., in][]{scot10} by taking the best-fit 
model parameters and using them to generate additional simulated sky 
realizations.  A \textit{P(D)} analysis was then carried out on each of 
these realizations to recover a set of new (scattered) best-fit model 
parameters.  This process was performed $\sim 100$ times with the same 
model inputs in order to generate a likelihood density map around the 
best-fit model parameters.

As an additional constraint on our model fitting and a means of
choosing between parametrizations, we also require
that the number counts model obey the constraint of the $1.2\rm\,mm$
CIB (see \S \ref{ss-CIB}), which is shown as the shaded region
in Figure \ref{f-PDerror}.  It is evident in Figure \ref{f-PDerror}
that although the power law parametrization can fit our observations well, 
it significantly overpredicts the $1.2\,\rm mm$ CIB.  In contrast, 
the Schechter function parametrization is in excellent agreement with the 
constraint of the CIB.  We therefore adopt the Schechter function 
parametrization as our fiducial model (e.g., Figure \ref{f-PDspace}) for the 
remainder of the paper.  At the highest flux densities 
the Schechter function and power-law models nominally predict very different 
behavior; however, our fluctuation analysis is not sensitive to the number 
counts at flux densities higher than those of our brightest detected sources.

The fact that both the power-law and Schechter function models of the 
differential number counts fit best when $S_{\rm cut}=0.05\,\rm mJy$ 
suggests that the 1.2\,mm number counts do not keep 
rising far beyond $0.05\,\rm mJy$;  formally, they may begin to fall between 
0.05 and 0.01\,mJy, or may already be decreasing by 0.05\,mJy.  The former
case is in agreement with recent surveys of lensing clusters 
\citep[e.g.,][]{knud08} that show SMG number counts increase at least 
as far down as $\simeq 0.1\,\rm mJy$ at 850\,${\rm \mu m}$.  At 1.2\,mm, 
this corresponds to $S_{1.2\,\rm mm}\simeq 0.03 - 0.04\,$mJy using the 
submillimeter spectral indices determined from matching detections at 
$850\,\mu\rm m$, $1.1\,\rm mm$, and $1.2\,\rm mm$ in the GOODS-N and COSMOS 
fields 
(Greve et al. 2008; Chapin et al. 2009; Austermann et al. 2010; see \S \ref{ss-compare counts})
This 
result is in contrast to the 
analysis of \cite{scot10}, who found that the choice of $S_{\rm cut}$ did 
not affect their results.  The discrepancy may be due to the fact that their lower
resolution ($28''$ HPBW) reduces the effective depth that can be reached before
sources begin crowding in the beam, thereby reducing the sensitivity of the $P(D)$
technique.

\section{Analysis of bright sources}

\subsection{Source extraction} \label{ss-ext}

We extracted sources from our ``best'' and ``full'' maps in a three-step 
process.  First, we minimized the chi-square statistic for a two-dimensional 
Gaussian profile with an $11^{\prime\prime}$ FWHM fit at each pixel center.  
This minimization was achieved quickly by using a matched filter convolution 
\citep[see, e.g.,][]{serj03}.  Given a signal image $S_{ij}$, an image of the 
local RMS noise $\sigma_{ij}$, and a smaller array describing the telescope's 
PSF $P_{xy}$, the chi-square statistic for a source with flux $F$ located at 
pixel $(i,j)$ can be written
\begin{equation}
\chi^2(F|i,j) = \sum_{xy}  \left(\frac {S_{i-x,j-y} - F P_{xy}}
{\sigma_{i-x,j-y}}\right)^2 
\end{equation}
We ignored the position-dependent negative sidelobes when we applied the 
matched filter and used only the central Gaussian profile, since the negative 
residual flux reaches only $\le 5\%$ of the peak positive intensity (see 
Figure \ref{f-negres}) in the map interior.  Additionally, our significant 
detections are on average farther away from each other than the largest chop 
throw used during the observations ($48^{\prime\prime}$), so 
their effect on 
our flux measurements will be less than 5\% of our strongest sources' flux 
densities (i.e., $\lesssim 0.25\,{\rm mJy}$).

\begin{figure} 
\epsscale{1.}
\plotone{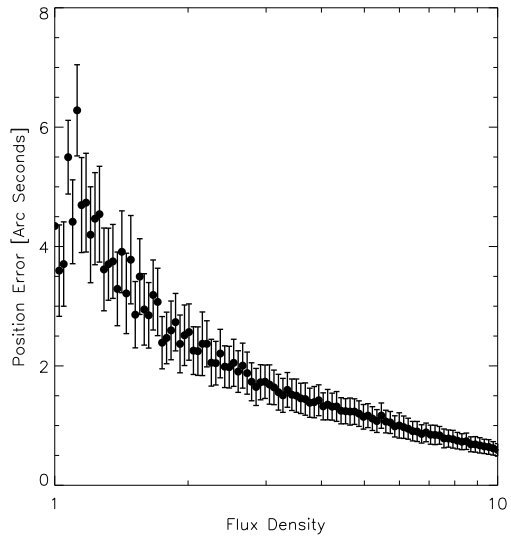}
\caption{Average angular separation between injected and recovered source 
locations as a function of flux density.  The injection process was 
performed 1000 times for each flux density value.  The scatter increases at 
low flux densities because of incompleteness.}
\label{f-posn}
\end{figure}

By finding the minimum of $\chi^2(F|i,j)$ as a function of $F$ and 
determining the associated uncertainty $\Delta F$ \citep{serj03}, we obtained 
\begin{equation}
\frac{d\chi^2}{dF}=0 \longrightarrow \frac{F}{\Delta F} = 
\frac{ \sum_{xy} 
S_{i-x,j-y} W_{i-x,j-y} P_{xy} } {\sqrt[]{\sum_{xy} W_{i-x,j-y} P_{xy}^2}}
\label{e-flux}
\end{equation}
as the S/N of each pixel, in terms of the weight 
map $W$ defined in \S \ref{ss-sigmap}.  Next, we located the centroid of each 
source to sub-pixel precision by fitting the PSF to the region in the original 
signal map at the location of each significant peak in the S/N map, allowing 
the position of the Gaussian to vary.  Figure \ref{f-posn} shows the 
uncertainty in this best-fit centroid position, derived via Monte Carlo 
simulations.  For the typical flux densities of our significant detections, 
the average offset between injected and recovered centroids is $1''-3''$.
Finally, we computed the best-fit flux density by taking the matched-filter 
weighted average of the flux map, this time with the PSF kernel shifted by 
interpolation to the more precise location of the source centroid.  After the 
flux density of each source was recorded, the source was removed from the map 
by subtracting the flux-scaled PSF from the source location before we searched 
for the next most significant detection.

By propagating the uncertainty in the signal image through
the $\chi^2$-minimization process,  \citet{serj03} have shown that the uncertainty 
in the resulting best-fit flux density at position $(i,j)$ is

\begin{equation}
\Delta F(i,j) =\frac{1}{\sqrt{\sum_{xy}\,W_{i-x,j-y}\,P_{x\,y}^{\,2}}} 
\end{equation}
We find that this expression consistently \textit{underestimates} the 
uncertainty in our map.  The reason is simply that the derivation by 
\citet{serj03} implicitly assumes that the Gaussian noise in the flux 
image is spatially uncorrelated.  In our images, the noise is correlated 
on a length exactly matching the FWHM of the telescope PSF, 
and pure noise fluctuations can be amplified along with the real point 
sources.  In order to correct for this underestimate we (1) produce an 
optimally filtered image and weight map, (2) use this filtered image 
and weight map to make a S/N ratio map, and (3) rescale the filtered weight 
map such that the standard deviation of the S/N map equals unity. This 
empirical calibration corrects for the effects of correlated noise 
and produces a map with accurate post-filtered flux density uncertainties.  
The accuracy of the method is evidenced by its matching the \textit{predicted} 
number of positive excursions in a correlated Gaussian field as a 
function of S/N level (see \S \ref{ss-spur}). 

The quoted $1\sigma$ errors (see columns $S_{\nu}^{\rm Best}$ and 
$S_{\nu}^{\rm Full}$ in Table \ref{t-sources}) for each detection correspond 
to the rescaled version of $\Delta F$ evaluated using the 
shuffled noise map at the location of the source (see \S \ref{ss-noimap}).  
This noise map describes the Gaussian noise of the observations more 
faithfully than the original signal map, which overestimates 
the noise by $\sim 5\%$ due to the positive and negative sidelobes from 
bright sources.  Because our noise is well above the estimated confusion 
limit (\S \ref{ss-conf}), Gaussian random fluctuations are the dominant 
source of uncertainty in our measurements.

\subsection{Comparison of results for best and full maps} \label{ss-hinge}

\begin{figure} 
\epsscale{1.}
\plotone{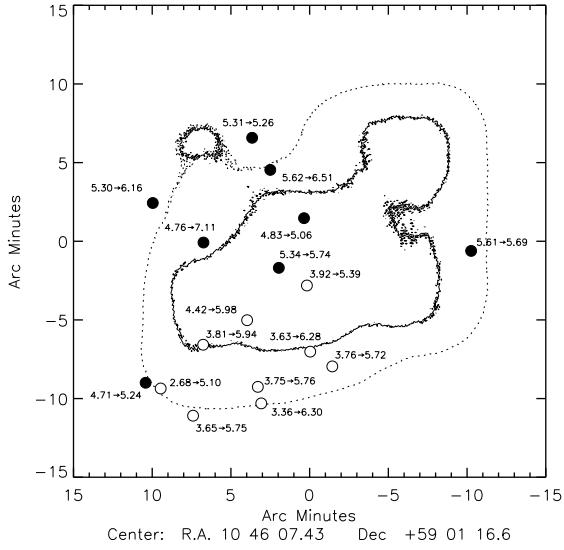}
\caption{The change in S/N of our 17 comparison sources in the ``best''
and ``full'' datasets.  Filled circles show eight sources in the ``best'' map 
with ${\rm S/N} \geq 4.5$.  In the ``full'' map, we recover all of these 
detections, plus others shown as empty circles, above a threshold ${\rm S/N} 
\geq 5.0$.  The solid/dashed contour denotes the $1\,\rm mJy\,beam^{-1}$ RMS 
noise threshold in the maps of the ``best''/``full'' data. The numbers 
near each detection show the S/N ratio of that source in the ``best'' 
data $\longrightarrow$ ``full'' data.}
\label{f-hinge}
\end{figure}

\begin{figure} 
\epsscale{1.}
\plotone{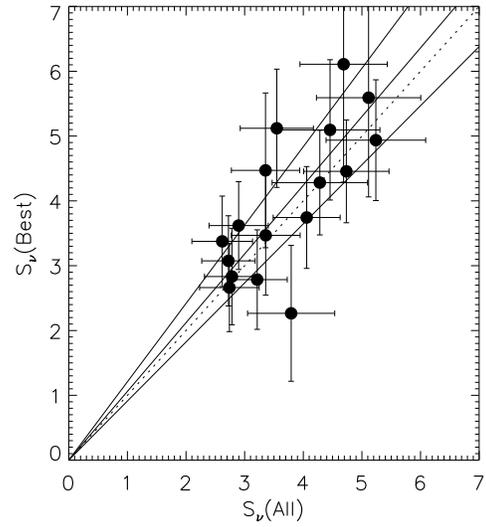}
\caption{Flux density comparison between the 17 highest S/N detections in 
the ``full'' data and their counterparts in the ``best'' data (these sources 
are plotted in Figure \ref{f-hinge}).  The solid lines show the best fit slope 
with $2\sigma$ uncertainties of a line constrained to cross the origin.  For 
illustration, the dotted line has unit slope. The chi-square minimization 
with \textit{x} and \textit{y} errors was performed using the IDL script 
\texttt{mpfit.pro} \citep{mark08}.
The best fitting slope is $m = 1.06\pm 0.07$ ($\pm 1\sigma$ uncertainties).}
\label{f-proof}
\end{figure}

As discussed in \S \ref{s-obs}, control software problems during our first 
two semesters of observations undermined our confidence in the reliability of 
the resulting maps.  To assess whether the ``full'' map could be trusted for 
bright source extraction, we performed two comparisons between our ``full''
data and observations with pristine calibration.  For the first 
comparison, we carried out the source extraction steps described 
in Section \ref{ss-ext} for {\it both} the ``best'' and the ``full'' datasets 
and compared the properties of the sources recovered from each.  Specifically, 
we began by choosing the eight sources with ${\rm S/N} \geq 4.5\sigma$ 
detections in our ``best'' map: above this threshold, we expect to see 
fewer than one spurious detection (\S \ref{ss-spur}).  All eight of these 
sources are recovered with $\geq 5.0\sigma$ significance in the ``full'' map.  
Figure \ref{f-hinge} shows the locations of these sources in the field, and 
the locations of the additional $\geq 5.0\sigma$ detections in the ``full'' 
map.  Each of the eight sources increases in significance between the ``best'' 
and the ``full'' maps.  Additionally, all but one of the ``full'' map's 17 
$\geq 5.0\sigma$ sources are identified in the ``best'' data at lower 
significance.  One source kept the same significance because it lies in the 
northeast corner of the field, where observations in the first two semesters 
contribute little additional sensitivity.  Further, for these 17 sources, 
the ratio of the flux densities in the ``best'' and ``full'' maps is 
consistent with unity (Figure \ref{f-proof}).

\begin{figure} 
\epsscale{1.}
\plotone{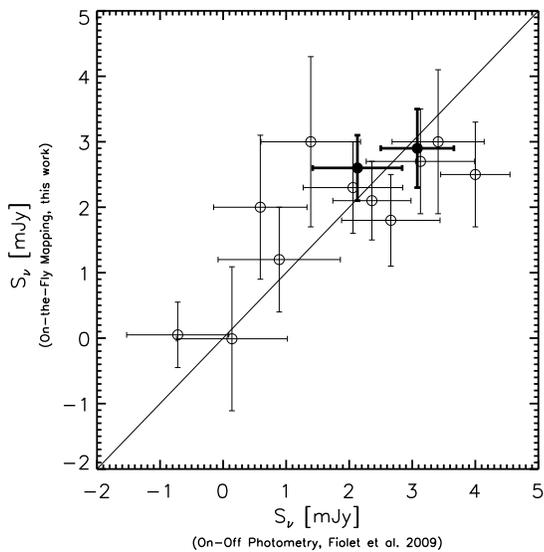}
\caption{Flux density comparison between on-off photometry-mode \citep{fiol09} 
and on-the-fly mapping (this work) of 13 \textit{Spitzer}-selected 
starburst galaxies in the LHN that lie within our ``full'' map coverage.  Symbols
with thick (thin) lines represent 2 (11) of the sources from \cite{fiol09} with 
significant detections (non-detections) in our full map.}
\label{f-fiolet}
\end{figure}

Next, we compared our ``full'' map to MAMBO on-off photometry by \cite{fiol09} 
of \textit{Spitzer}-selected high-redshift starburst candidates in the LHN.  
We tabulated the map flux densities at the positions of the 13 galaxies 
in their sample that lie within our map's footprint (two of these turn 
up as significant detections in our ``full'' map; we use these sources' 
non-deboosted 
flux densities here) and 
compared them to the flux densities reported by \cite{fiol09}.   We found that
the flux densities from the two significant detections as well as those from 10
of the 11 non-detections are consistent to within $1\sigma$ (see Figure 
\ref{f-fiolet}; one source is only consistent to within $\sim 1.5\sigma$).

These two successful consistency checks lead 
us to conclude that the errors in our first two semesters' data are not at a 
level that compromises point source detection, at least for high-significance 
sources.  We have therefore proceeded to define our source catalog on 
the basis of the ``full'' map.  Since the fluctuation analysis decribed in 
\S \ref{s-pd} relies on the 
authentic reproduction of the field's noise properties and low-S/N 
fluctuations, we have restricted this analysis to the ``best'' data only.

\subsection{Confusion} \label{ss-conf}
Random Gaussian noise is the uncertainty in the total flux density inside any 
single \textit{beam} on the sky due to random fluctuations, while confusion 
noise is an additional uncertainty in the flux density of a single \textit{source} 
due to the contributions of faint sources within that beam.  The 
``confusion limit'' is defined as the flux density threshold at which 
confusion noise significantly affects the measured flux density 
of a source, and is commonly taken to be the flux density 
above which the integrated number counts of all brighter sources reach $\simeq 0.033$ per beam
\citep{cond74}.  
In our map, this rule gives $\simeq 0.9\,\rm mJy$ using $\theta_B = 15.6^{\prime\prime}$ 
(in the smoothed $\leftrightarrow$ filtered version of our ``full'' map) and 
assuming our best-fit number counts (\S \ref{s-pd}).  

We have also made a direct estimate of the confusion noise by comparing 
the noise in the central regions of the filtered ``best'' map to the same 
region in a series of filtered jackknifed noise maps.  Since the jackknifed 
noise maps remove confused as well as bright sources, the increase in average 
RMS noise in this relatively uniform region indicates our map contains confusion 
noise at the level of $\sigma_C \simeq 0.24\,\rm mJy\, beam^{-1}$.
As a consistency check, we  have also estimated the confusion noise by 
generating simulated maps with source populations following our best-fit 
number counts from 0.05\,mJy up to the confusion limit of $0.9\,\rm mJy$.  
Due to the central limit theorem, these faint and confused maps with zero 
mean have roughly Gaussian PFDs and provide approximations of the confusion 
noise, assuming our model of the number counts.  The standard deviation 
in these maps is $0.21\,\rm mJy\,beam^{-1}$, in agreement with the measured
confusion noise within the uncertainties of the number counts model.
We therefore adopt the measured value of $\sigma_{C}\simeq 0.24\,\rm mJy$
as our estimate of the confusion noise.  The average uncertainty in the 
flux density of a source in our catalog is $0.62\,{\rm mJy}$, indicating 
that confusion does not dominate our noise budget.

\subsection{Spurious sources} \label{ss-spur}

\begin{figure} 
\epsscale{1.}
\plotone{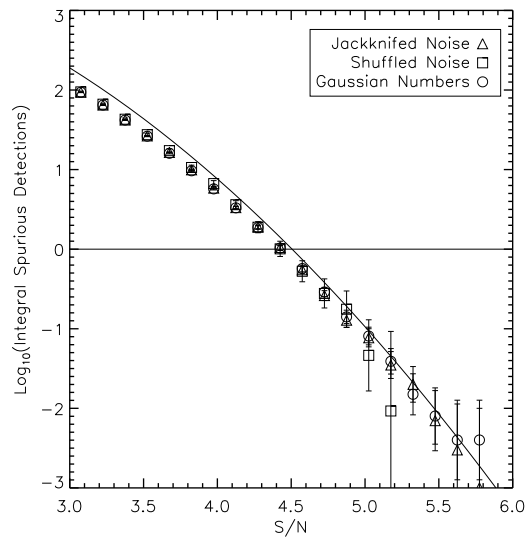}
\caption{Integral numbers of spurious detections as a function of S/N within
jackknifed noise maps, shuffled noise maps, and maps of random Gaussian 
numbers.  We 
expect to detect 0.8 spurious sources with ${\rm S/N} \geq 4.5$ and $\simeq 
5.4$ spurious sources with ${\rm S/N} \geq 4.0$.  The curve shows the expected 
number of excursions above a given S/N level within a Gaussian random field 
in the limit of high S/N \citep{adler81}.}
\label{f-spur}
\end{figure}

We estimated the number of spurious detections as a function of S/N by running 
our source extraction algorithm on various noise maps (see \S \ref{ss-noimap}).
We tested jackknifed noise maps, shuffled noise maps, and simple Gaussian 
random numbers.  The Gaussian random numbers had a spatially varying standard 
deviation matched to the weight map of the observations.  Figure \ref{f-spur} 
shows the mean total numbers of spurious detections found in $10^3$ jackknifed 
and Gaussian number noise maps, and in $10^2$ shuffled noise maps, as a 
function of S/N.  All three styles of noise map are consistent with each other 
in their ability to produce spurious detections with $\rm S/N \ge 3.0$.  This 
result confirms that for the purposes of extracting high-significance 
detections, the shuffled noise maps are just as ``source-free'' as the 
jackknifed noise maps.  Additionally, both are consistent with a Gaussian 
distribution down to $3.0\sigma$ (and likely consistent with Gaussian noise 
at all S/N, as implied by the PFD histograms in \S \ref{f-histo}).  The 
over-plotted curve in Figure \ref{f-spur} shows the expected number of 
excursions above a given S/N level in any isotropic and homogenous Gaussian 
random field, derived (and thus only formally valid) for high excursions 
\citep[see, e.g., Chapter 6 of][]{adler81}.  The agreement at high S/N 
indicates that our noise maps and source extraction algorithm are 
well-behaved.  For both the ``best'' and ``full'' maps, we expect 0.8 (5.4) 
spurious sources with $\rm S/N \ge 4.5\,(4.0)$.

A source at high risk of being a spurious detection can be 
identified by calculating the total probability that the deboosted flux
density is $\le 0\,\rm mJy$ \citep[see, e.g.,][]{aust10,scot10}, hereafter 
referred to as $P(S<0)$.  Using the threshold of $P(S<0)\ge0.10$ used by 
\citet{aust10}, we identify only one (ID \# 33) high-risk spurious 
detection in our $\rm S/N>4.0$ sample (see Table \ref{t-sources}).  

\subsection{Completeness} \label{ss-comp}

\begin{figure} 
\epsscale{1.}
\plotone{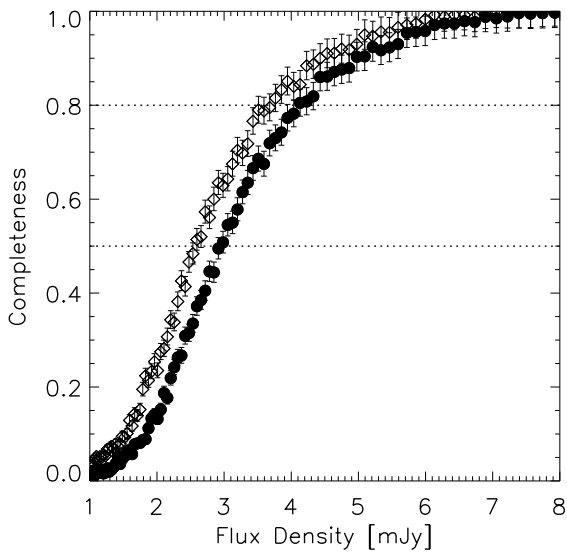}
\caption{Completeness as a function of flux density.  The 
circles/diamonds represent $4.5\sigma/4.0\sigma$
source extraction thresholds for detection.  The number count calculation
uses a threshold of $4.5\sigma$, while our source catalog includes sources
down to $4.0\sigma$.  Horizontal lines represent 80\% and 50\% completeness
limits.}
\label{f-comp}
\end{figure}

We estimate the completeness in our data using the Monte Carlo method of 
searching for injected sources of varying flux density.  The inhomogeneous 
noise in our map means that sources of identical flux density have different 
probabilities of being detected in different locations.  We account for this 
effect statistically by performing the completeness simulation assuming sources 
have equal probability of being located anywhere in the map during the 
injection process.  Although the high-redshift star-forming galaxy population 
that our observations trace is likely to exhibit clustering, with the 
brightest galaxy mergers occurring in the most massive dark matter halos 
\citep[see, e.g.,][]{weiss09}, the large $\Delta z$ interval to which 
millimeter selection is sensitive tends to weaken angular clustering 
signatures (see however \S \ref{ss-clustering}).  We inject model sources with 
known flux density into our original signal map one at a time at random positions 
and search for them using the same source extraction algorithm used to create 
our source list.  If an artificial source is recovered with ${\rm S/N} \geq 
4.0$ within $11''$ of the injected location, it is considered detected.  The 
injection process was repeated $10^3$ times for each flux density in a 
logarithmic 
grid from 1.0\,mJy to 10.0\,mJy; the average recovery 
percentage is shown in Figure \ref{f-comp}.  Our map is 80\% complete
at $3.7\,\rm mJy$ and 50\% complete at $2.6\,\rm mJy$.
We also tabulated the angular separations 
between the injected and recovered source positions to characterize the 
uncertainties in the positions of our actual significant detections (ignoring 
telescope pointing errors).  For $S_{\rm 1.2\,mm} 
\gtrsim 2\,\rm mJy$, the average position error $\langle \Delta \theta 
\rangle \leq 3''$ (see Figure \ref{f-posn}).

\subsection{Flux boosting} \label{ss-boost}

\begin{figure} 
\epsscale{1.}
\plotone{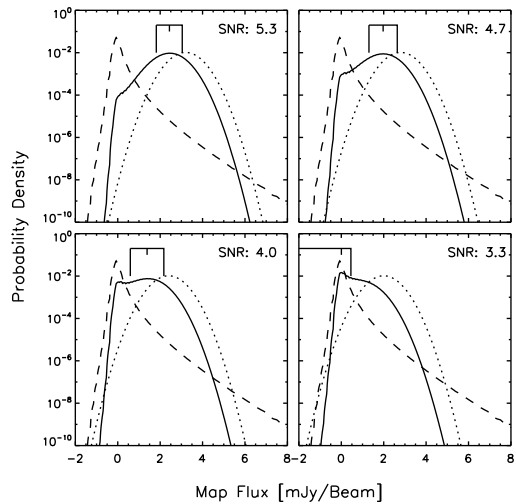}
\caption{Posterior probability distributions of four characteristic 
S/N regimes.  The dotted lines represent the Gaussian probability 
distributions for each of the four measurements (the likelihoods), all 
assuming $\sigma=\,0.6 \rm \, mJy$ and varying mean. The dashed lines 
represent the prior flux distribution constructed from Monte Carlo 
simulations, and the solid lines represent the the normalized posterior 
probability distributions.  The brackets above the curves denote the 
peak likelihood values (the deboosted flux densities) 
along with the left and right 
38$\%$ confidence intervals.  When the S/N is too low for either the 
left or the right confidence interval to converge, as is the case in the 
final panel, we use instead the analytic formula from Equation \ref{e-deboost}.}
\label{f-pfd}
\end{figure}

To correct the measured flux densities of our detections for 
the effect of
``flux boosting,'' we  use the Bayesian technique described in 
\cite{coppin05}, which because of its versatility in handling both chopped and 
unchopped data has been adapted for use at many wavelengths 
\cite[e.g.,][]{coppin06,grev08,scot08,perer08,scot10,aust10}.
Using Bayes's Theorem and the prior information of the number counts 
functional form found from our $P(D)$ analysis (\S \ref{s-pd}), the 
probability that a source has true flux density $S_0$ given a measurement 
$S$ with uncertainty $\sigma$ is equal to
\begin{equation}
P(S_0|S,\sigma) = \frac{P(S,\sigma| S_0) P(S_0)}{P(S,\sigma)}
\end{equation}
where $P(S,\sigma|S_0)$ is the posterior probability, $P(S_0)$ is the prior 
flux density distribution, $P(S,\sigma| S_0)$ is the likelihood, and $P(S, 
\sigma)$ is the prior measurement distribution.  $P(S, \sigma)$ is independent 
of $S_0$, so only acts to normalize the expression such that $\int P(S_0|
S,\sigma) dS_0 = 1$; hereafter, it will be ignored.  We have shown that the 
uncertainty in our map is dominated by Gaussian random noise, so $P(S,\sigma| 
S_0)$ takes the form 
\begin{equation}
P(S,\sigma| S_0) \propto {\rm exp}\left[-\frac{\left(S-S_0\right)^2}
{2\sigma^2}\right]
\end{equation}
To estimate the prior flux distribution in the map ($P(S_0)$), we assembled a 
PFD containing the pixels from $10^4$ noise-free random sky realizations 
that used the best-fit number count parameters from our \textit{P(D)} analysis.
The peak value and 68\% double-sided confidence intervals of the resulting 
posterior probability function ($P(S_0|S,\sigma)$) were found numerically for 
each measured flux density and uncertainty.  Figure \ref{f-pfd} shows four 
examples of the deboosting process, each in a different regime of source S/N.  
If the S/N is too low, and the integration of the confidence 
intervals does not converge, we instead use an analytic formula to estimate the
deboosted flux density.  For this, we generalize the formalism of \cite{hogg98} 
to a Schechter function, and locate the maximum of the posterior flux distribution:

\begin{equation}
P(S_0|S,\sigma) \propto S_{0}^{-\delta'}\,\exp\left[ -\frac{S_{0}}{S_{\rm exp}'}-
 \frac{\left(S-S_0\right)^2}{2\sigma^2}\right]
\end{equation}
where $\delta'$ and $S_{\rm exp}'$ are the power-law slope and exponential
scale factor of the Schechter function, respectively.  By solving for 
$S_{0}$ when the derivative of the above 
expression vanishes, we find the highest posterior probability to be
achieved for

\begin{equation}
S_{\rm true} = \frac{S \, S_{\rm exp}'-\sigma^2 + \sqrt{
\left(\sigma^2-S \, S_{\rm exp}'\right)^2 -4 \,\delta'\,S_{\rm exp}'^2\sigma^2 }}{2 S_{\rm exp}'}
\label{e-deboost}
\end{equation}
 
\begin{figure} 
\epsscale{0.8}
\plotone{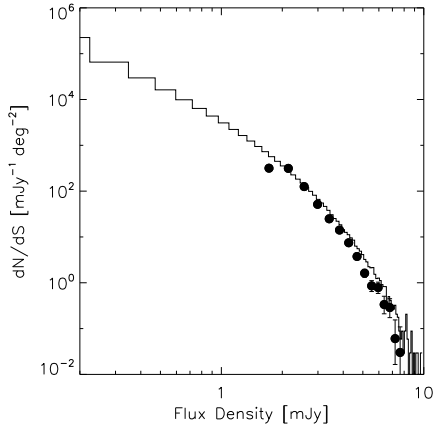}
\caption{Monte Carlo simulation to test our Bayesian flux boosting 
correction.  The histogram shows the average number counts model 
used to inject sources into realistic noise maps.  The circles
show the average number counts 
calculated from the raw counts by using the Bayesian method to 
deboost the flux density of each recovered source.  Error bars 
represent only the Poisson uncertainty in the average \citep{gehr86}.}
\label{f-numbertest}
\end{figure}

To ensure that our adaption of the Bayesian method of flux 
deboosting returns a properly corrected estimate of the true 
number counts, we performed a Monte Carlo simulation to directly 
calculate the observed number counts of random sky realizations populated 
with source distributions following our best-fit number counts 
\cite[see, e.g.,][]{coppin06}.  Figure \ref{f-numbertest} shows 
the results of this consistency check.  
This simulation demonstrates that the Bayesian method of flux deboosting 
performs well in recovering the original injected number counts.  
The residual scatter of the average recovered 
number counts around the average input model in Figure \ref{f-numbertest} demonstrates the level of systematic error
in the algorithm, which is significantly smaller than the statistical
error of our differential number counts estimate 
(see Figure \ref{f-counts}).

\subsection{Direct calculation of number counts} \label{ss-counts}

While our catalog of detections includes all sources with ${\rm S/N} > 4.0$, 
we use only detections with ${\rm S/N} > 4.5$ for our direct calculation of 
the number counts because above this threshold, we expect to detect less than 
one spurious source (see Figure \ref{f-spur}).  Table \ref{t-counts} presents 
integral and differential number counts after correction for completeness
and flux boosting.  Figure \ref{f-counts} shows our directly 
calculated number counts, along with 
the 95\% confidence regions for the best-fit power law and Schechter function
models of the differential number counts found from the 
\textit{P(D)} analysis.  These two independent methods of estimating the 
number counts are in agreement with each other.  This consistency is
encouraging because the \textit{P(D)} analysis and the direct estimate of 
number counts depend on the faint and bright pixel values in different ways. 

\subsection{Clustering} \label{ss-clustering}

The group of sources in the southeast corner of our field, as well as the 
large void in the center, prompted us to perform a clustering analysis to 
determine whether or not the distribution of sources in our map is 
statistically clustered or not. To perform the analysis, we used the 
\cite{landy93} correlation function estimator:
\begin{equation}
w(\theta)=\frac{DD-2DR+RR}{RR}
\end{equation}
with variance
\begin{equation}
\langle \Delta w(\theta) \rangle ^2 \simeq \frac{(1+w(\theta))^2}{RR}
\end{equation}
\citep{gawi06}.  
In the equations above, \textit{DD, RR}, and \textit{DR} represent 
the normalized numbers of unique galaxy-galaxy, random-random, and 
galaxy-random pairs with angular separations $\theta \pm d\theta/2$.  
This estimator is used frequently in extragalactic deep field analyses 
\citep[e.g.,][]{borys03,scot06,weiss09}, and has been shown to have 
nearly Poisson variance and zero bias \citep{landy93}.  We take into 
account the geometric boundary of the map and the variation in 
detectability with position by generating the random locations with 
the same Monte Carlo algorithm used for the \textit{P(D)} analysis.  
We inject ensembles of sources following our best fitting number counts into 
a noise map at random locations and use the positions of sources 
detected with ${\rm S/N} >4.0$ as our random coordinates.  This Monte Carlo 
technique is important for ensuring that we do not misinterpret 
depth variation in the map as a clustering signal.  

\begin{figure} 
\epsscale{0.7}
\plotone{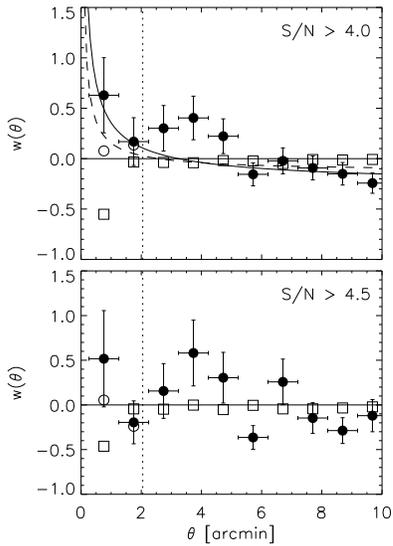}
\caption{Landy-Szalay correlation function estimator $w(\theta)$ as a 
function of angular separation.  Top panel uses our 41 detections with 
${\rm S/N}>4.0$.  Bottom panel uses 27 sources with ${\rm S/N}>4.5$.  The 
vertical dashed line shows upper limit in $\theta$ on the clustering 
suppression effect introduced by chopping (see \S \ref{ss-clustering}).  Open squares show the results from using
random positions to check the zero-clustering baseline; open circles show the 
raw clustering signal, which are corrected for the zero-clustering baseline
to deliver the filled circles.}
\label{f-LScorr}
\end{figure}

To confirm that this technique is unbiased, we also performed the full 
clustering analysis on only random positions to check that we recovered a 
flat $w(\theta)=0$ response (see Figure \ref{f-LScorr}).  For small 
separations ($\lesssim 2'$), however, it turns out that $w(\theta)$ does 
not return zero in our data: depending on the position within 
the map, the negative sidelobes can suppress the flux densities of 
nearby sources enough to lower their S/N ratios below the detection 
threshold.  This effect begins to have an effect at $\simeq 2\times$ 
the chop throw (of which the maximum used in any semester was $48''$),
and has a strong effect at separations $\leq 1\times$ chop throw.  
Because this effect suppresses the detection of \textit{RR} and 
\textit{DD} pairs but not \textit{DR} pairs, the zero-clustering 
baseline for chopped data like ours is less than zero at these small 
angles.  In order to assess the clustering in the map while taking 
this bias into account, we measure the effective clustering relative 
to the zero-clustering baseline for these separations ($\lesssim 2'$).  

The result of our Monte Carlo clustering analysis is shown in Figure 
\ref{f-LScorr}.  We find a small clustering signal when using all 
detections with ${\rm S/N} >4.0$ that agrees reasonably well with 
the angular correlation function measured by \cite{scot06}, who 
combined many different SCUBA $850\,{\rm \mu m}$ blank field maps, 
and that shows stronger clustering (albeit at lower S/N) than the 
correlation function measured by \cite{weiss09} at $870\,\mu\rm m$ in 
the ECDF-S. \citet{will11} have analyzed the clustering of
3.0--$3.5\sigma$ $1.1\,\rm mm$ detections in a $0.72\,\rm deg^2$ map 
of the COSMOS field with ASTE/AzTEC, concluding that it is
difficult to recover reliable clustering parameters for SMGs 
from maps whose angular resolution and total area are limited. 
This result argues for caution in interpreting our clustering 
analysis, although we do benefit to an extent from MAMBO's 
relatively high angular resolution.
An interesting feature in our correlation function is the 
spike near $\theta\simeq4'$.  This signal is due to the rich 
group of sources in the southeastern corner of the map, all at typical 
relative spacings of a few arcminutes from each other (see also \S 
\ref{ss-radiocorr}).  It may be noteworthy that \cite{weiss09} find a
$\sim 2.4\,\sigma$ spike above their best fitting model of 
$\omega(\theta)$ at a scale of $\sim 5'$, near where \citet{will11}
also detect a slight positive excess in $\omega(\theta)$.  When performing 
the analysis on only our (27) most significant sources with 
${\rm S/N} >4.5$, we find no significant clustering signal.

\begin{figure} 
\epsscale{1.}
\plotone{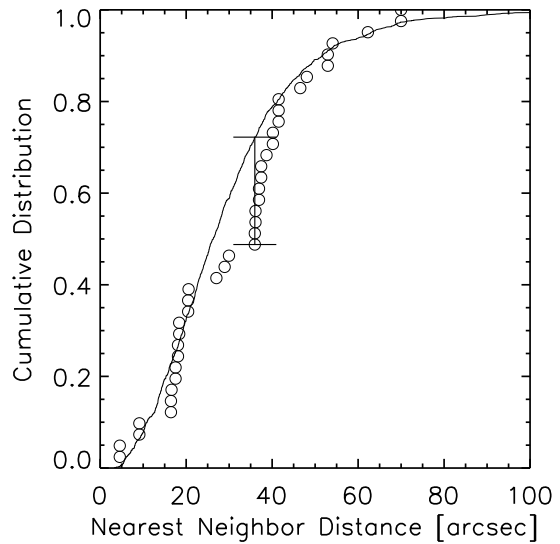}
\caption{Integral distribution of nearest neighbors.  Solid line shows the distribution
from Monte-Carlo generated random positions.  Circles show the distribution of our 41
significant detections.  The vertical line segment denotes the maximum difference
between the two distributions.  A Kolmogorov-Smirnov test rules out the hypothesis that
our sources are drawn from a random distribution at 95\% confidence.}
\label{f-NNcorr}
\end{figure}

Following the analysis of, e.g., \citet{borys03} for the SCUBA 
``Supermap,'' we also use the method of \cite{scott89} to 
analyze the cumulative distribution of nearest neighbors to test 
whether our galaxy positions are consistent with being drawn from a 
random distribution (see Figure \ref{f-NNcorr}).  Because the nearest 
neighbor analysis is sensitive to the total number of positions 
used, we use the 41 most significant detections in {\it each} Monte Carlo 
realization, instead of all of those detections with ${\rm S/N}>4.0$ as 
in the correlation function analysis.  Because the number counts 
rise quickly, the S/N of the least significant discovered source 
varies, but is always close to 4.0.  A Kolmogorov-Smirnov test rules 
out the null hypothesis that our significant detections are 
drawn from a random position distribution at the 95\% confidence 
level, implying that the source locations in the map (e.g., defining the 
southeastern clump and the central void) are not arranged randomly.

\section{Counterpart identification} \label{ss-counterpt}

We have calculated the corrected probability of chance associations
\citep[$P$;][]{down86} between our MAMBO detections and possible counterparts at the
other wavelengths at which the LHN has been observed 
(see Table \ref{t-sources}).  The $P$ statistic is defined by

\begin{equation}
P=1-e^{-E}
\end{equation}
for $E=P^*[1+{\rm ln}(P_c/P^*)]$, $P^*=\pi r^2 N(>S)$, and 
$P_c=\pi r_s^2 N_c$, in terms of the brightness of the counterpart $S$, 
the source separation $r$, the search radius $r_s$, the
number density of sources brighter than $S$ $N(>S)$, and  
the number density of sources at the sensitivity limit $N_c$.
Based on the results of the position error 
analysis (see Figure \ref{f-posn}), we chose a counterpart search radius 
of $8^{\prime\prime}$.  Because positional uncertainty 
$\rm \sigma \propto FWHM \times SNR^{-1}$, our $11''$ beam is the dominant 
source of error, and we ignore the positional uncertainties at other 
wavelengths.  We consider $P < 0.01$ to define a robust counterpart, 
$0.01 \le P < 0.05$ a likely counterpart, and $P \geq 0.05$ an unlikely 
association.

\subsection{20\,cm counterparts} \label{ss-20cm}

We used the NRAO VLA Sky Survey (NVSS) and the deep SWIRE number counts of 
\citet{cond98} and \cite{owen08}, respectively, in the calculation of 
\textit{P} to assess the significance of 20\,cm counterparts.  The 
20\,cm VLA pointing of the LHN has a central RMS sensitivity of $2.7\,{\rm \mu 
Jy}$, rising to $\sim 4$--$5\,\rm\mu Jy$ near the edges of 
our MAMBO map.  When we compare our 41 ${\rm S/N}>4.0$ detections to the 
$5\sigma$ 20\,cm catalog of \cite{owen08}, 44\% (18) have robust 
counterparts, and 41\% (17) have likely counterparts.  We have also 
reexamined the $20\,\rm cm$ map in the vicinity of the remaining MAMBO 
sources and have identified one additional robust counterpart (ID\,\#\,9), 
two likely counterparts (ID\,\#\,28 and \#\,36), and one unlikely counterpart (ID\,\#\,20) 
at the 4--5$\sigma$ level.  We also deblended one likely counterpart into
one robust and one unlikely counterpart (ID\,\#\,17).  After including these 
additional sources, $49\%\,(20)$ of our MAMBO detections have robust 
counterparts, $44\%\,(18)$ have likely counterparts, and $7\%\,(3)$ have 
unlikely or no detected counterparts.  We performed a Monte Carlo simulation 
to test the reliability of our $P$ values and found that $4.9\pm 0.2\%$ of 
randomly chosen positions within our MAMBO field have a likely radio 
counterpart ($P < 0.05$) within $8''$, confirming the validity of the high 
number of robust associations.  We expect $\sim 5$ spurious detections 
above ${\rm S/N} > 4.0$; thus, the handful of sources with unlikely or 
no radio counterparts may be spurious detections if they do not lie at a
very high redshift.  One MAMBO source (ID\,\#\,32) with an unlikely 
($P_{20\,\rm cm}=0.056$) radio counterpart also has a likely 
($P_{250\,\rm\mu m}=0.016$) \textit{Herschel} counterpart \citep{magd10}, 
arguing against its being a spurious detection.

\subsection{50\,cm radio counterparts} \label{ss-50cm}

We have extracted $50\,\rm cm$ flux densities from the GMRT map \citep{owen11a} 
with the same technique used at 20\,cm and 50\,cm \citep[see][]{owen08,owen09b}. 
The uncertainties listed in Table \ref{t-sources} reflect the local RMS noise
in the image and do not include a $\sim3\%$ calibration error or a spatially
varying GMRT pointing error.  Two of the 50\,cm detections are heavily blended
with bright neighbors, so for these counterparts we report only tentative 
fluxes.  Of the 40 tabulated $20\,\rm cm$ counterparts (including the two 
with $P>0.05$), all 40 have 50\,cm counterparts.  The one 20\,cm non-detection 
(within $8''$) is also a 50\,cm non-detection.

\subsection{90\,cm radio counterparts}

To search for 90\,cm counterparts, we used the 90\,cm radio 
catalog of \cite{owen09b}, which has an RMS sensitivity of $10\,\rm\mu Jy$.  Of our 41 
MAMBO sources, $24\%$\,(10) have 90\,cm counterparts.  Each of the ten 
90\,cm counterparts is also detected at 50 and 20\,cm with $P_{20\,\rm cm}<0.05$.

\subsection{$\mathbf{24\,\mu}$m counterparts}

In addition to SWIRE $24\,\rm\mu m$ observations of the LHN ($3\sigma$ 
depth of $209\,\mu\rm Jy$), there exist deeper {\it Spitzer}/MIPS data with a
$3\sigma$ depth of $18\,\rm\mu Jy$ \citep{owen11b}.  We searched for 
$24\,\rm\mu m $ counterparts to our MAMBO detections in this deeper MIPS 
image.  To calculate \textit{P} statistics for $24\,{\rm \mu m}$ counterparts, 
we used the counts of \citet{beth10}.  A Monte Carlo simulation of the 
$24\,\rm\mu m$ \textit{P}-statistic finds that $4.7\pm0.3\%$ of random 
positions yield a counterpart with $P<0.05$.  Within our sample of 41 MAMBO sources, 
20$\%$\,(8) have robust $24\,{\rm \mu m}$ counterparts, and 29\,$\%$\,(12) have 
likely counterparts.

\subsection{X-ray counterparts} \label{ss-xray}

Only one source (MM\,J104522.8+585558 = ID \# 26) has a likely X-ray 
counterpart \citep[CXOSW\,J104523.6+585601;][]{wilk09}.  The X-ray source is at a distance 
of $7.2''$ and has a broad band (0.3--8.0\,keV) flux of $(2.5 \pm 1.1) \times 
10^{-15}\,\rm erg\,cm^{-2}\,s^{-1}$ \citep{poll06}.  By using the 
2.5--$7\,\rm keV$ flux of $1.58\times 10^{-15}\,\rm erg\,cm^{-2}\,s^{-1}$ 
together with the \textit{Chandra}/SWIRE counts from 2--$8\,\rm keV$ 
\citep{wilk09} we can set an upper limit on the probability of 
chance association of $P \lesssim 0.02$.

\section{Discussion}

\subsection{Number counts vs. previous deep fields} \label{ss-compare counts}

\begin{figure} 
\epsscale{1.}
\plotone{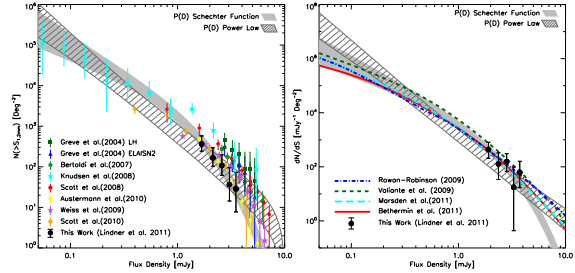}
\caption{\textit{Left}: 1.2\,mm integral number counts, compared to the 
\textit{P(D)} analysis best fit models and other observed number counts 
from the literature.  Filled black circles show our Bayesian 
deboosted number counts with 95\% confidence error bars including the 
Poisson uncertainty and the uncertainty in the 
completeness correction.  The 
solid filled region shows the 95\% confidence region for the best fitting 
Schechter function from the \textit{P(D)} analysis.  The cross-hatched region shows 
the same for the best fitting power law.  MAMBO counts: LH and 
ELAIS-N2 \citep[green squares and blue triangles;][]{grev04}, 
COSMOS \citep[green circles;][]{bert07}.  LABOCA counts: ECDF-S 
\citep[purple stars;][]{weiss09}. AzTEC counts: GOODS-S 
\citep[orange circles;][]{scot10}, SHADES \citep[yellow circles;][]{aust10}, 
COSMOS \citep[red circles;][]{scot08}.  SCUBA lensing cluster counts from 
\cite{knud08} are shown as red stars.  
  The counts at $850\,\rm \mu m$ (SCUBA), 
$870\,\rm\mu m$ (LABOCA), and $1.1\,\rm mm$ (AzTEC) have been scaled to 
$1.2\,\rm mm$ (see \S \ref{ss-compare counts}).  \textit{Right}: 1.2\,mm 
differential number counts and \textit{P(D)} models compared to the predictions 
of galaxy evolution models.  Lines represent different 
differential counts predictions 
by \cite{rowa09}, blue dot-dashed; \cite{vali09}, green short-dashed; 
\cite{beth11}, red solid; \cite{mars11}, cyan long-dashed.  Models with 
predictions only for the $1.1\,\rm mm$ waveband were scaled in order 
to compare to our observations (see \S \ref{ss-compare counts}).}
\label{f-counts}
\end{figure}

Previous deep surveys at 1.2\,mm using MAMBO \citep[e.g.,][]{grev04,bert07} 
have returned directly calculated 1.2\,mm number counts in the Lockman Hole
East (LHE), ELAIS-N2, and COSMOS fields.  The parameters of these surveys 
are listed in Table \ref{t-areas}; we compare their results to our 
directly calculated counts, as well as to our best-fit $P(D)$ models, in 
Figure \ref{f-counts}.  We find that our power-law slope is consistent with 
their results, but our results
have a lower overall normalization.  This difference in normalization 
might be due to the different methods used in the number counts 
calculations.  We have used the Bayesian method of flux 
deboosting presented in \cite{coppin05}, and only include our most 
significant detections in the calculation.  The analyses of 
\cite{grev04} and \cite{bert07} use the method of injecting sources 
into noise maps to determine their flux deboosting correction, and 
include sources with lower S/N in their number counts calculation.  
In principle, any S/N cutoff would be acceptable for the latter calculation 
as long as the completeness correction uses the same threshold; however, 
lower S/N thresholds will lead to more spurious detections.  Both of these 
effects could be contributing to their higher normalization.  
However, considering the relatively large error bars on all the  
measurements and the internal variation among the ELAIS-N2, LHE, 
and COSMOS datasets themselves, the results are still nearly consistent.

Because a single power-law parametrization of the
number counts is commonly used to compare the results of deep surveys, we 
begin by noting that
our best-fit power law index ($\delta=3.14^{+0.14}_{-0.18}$) is consistent 
with the results of surveys at other wavelengths that fit their number counts 
using a similar (single power-law) model.  \cite{coppin06} find $850\,\mu \rm 
m$ power-law indices of $\delta=2.9\pm0.2$ and $\delta=3.0\pm0.3$ in the LHE
and Subaru/XMM-Newton deep fields, respectively.  Using Bolocam data at 
1.1\,mm, \cite{laur05} estimate a power-law index of $\delta=3.16$ from 
directly calculated counts in the Lockman Hole.  However, \cite{malo05} 
performed a \textit{P(D)} analysis on the same 1.1\,mm Bolocam data and 
find $\delta=2.7^{+0.18}_{-0.15}$.  Although it is well within the $1\,\sigma$ 
uncertainties of the \cite{laur05} result, the latter slope differs from ours 
by $>2\sigma$.  In this case, differences in the methods of our \textit{P(D)} 
analyses might be the differentiating factor.  \cite{malo05} 
also used chopped observations in their analysis, for example, but ignored 
the effects of chopping on the PFD.  It is possible 
that by not including the negative residual flux in their \textit{P(D)} 
analysis, they required many fewer faint sources to match the pixel 
distribution of the real data (and therefore derived a shallower power law 
slope).  However, a value of $\delta \simeq 2.7$ is also the best-fit single 
power-law slope found by  \cite{scot10} for their \textit{P(D)} 
analysis of unchopped 1.1\,mm AzTEC data in the GOODS-S field.

In Figure \ref{f-counts} (see also Table \ref{t-areas}), we also compare 
our results to those for deep 
field observations at 1.1\,mm by AzTEC  of the COSMOS \citep{scot08}, 
GOODS-N \citep{perer08}, GOODS-S \citep{scot10}, SHADES \citep{aust10}, 
and AKARI, SSA-N2, and SXDF \citep{hats10} deep fields.  (For clarity, the
observations of \cite{hats10} and \cite{perer08} are not shown in the plot
because their data points lie within the scatter of the other AzTEC
observations.)  We also 
compare our results to the extremely deep SMG counts measured in 
lensing fields at $850\,\mu\rm m$ by \cite{knud08} as well as the 
recent wide map by \cite{weiss09} using LABOCA 
at $870\,\mu\rm m$ in the ECDF-S.
In order to compare our number counts directly to the results of these surveys 
at other wavelengths, we rescale their flux densities.
Our choice of rescaling factor is based on the direct 
comparisons between $S_{850 \rm \mu m}$, $S_{1.1\rm mm}$, and 
$S_{1.2\rm mm}$ for galaxies in the GOODS-N field.  The average 
flux density  ratio for sources with robust SCUBA and AzTEC detections
in the GOODS-N field is 
$S_{850\,\rm \mu m}/S_{1.1\,\rm mm}\simeq 1.8 -2.0$ \citep{perer08,chapi09}.  
When comparing SCUBA and MAMBO detections, \cite{grev08} find 
$S_{850\rm \mu m}$/$S_{1.2 \rm mm}\simeq 2.5$.  By coadding the MAMBO 
and AzTEC observations in the GOODS-N
field into a map at an effective wavelength of $\lambda =1.16\,\rm mm$,
 \citet{penn11} find an average value of $S_{1.16\rm mm}$/$S_{1.1 \rm mm}\sim 0.88$
and $S_{1.16\rm mm}$/$S_{1.2 \rm mm}\sim 1.14$.  All of these 
results are consistent with a single modified blackbody spectrum, for 
$\beta=1.5$ and $T_{d}=30\,\rm K$, observed at $z\simeq 2.5$.  Therefore, we 
adopt this fiducial 
galaxy model when comparing fluxes at different wavelengths and use 
$S_{850\,\rm \mu m}/S_{1.2\,\rm mm}=2.3$, $S_{870\,\rm \mu m}/S_{1.2\,\rm 
mm}=2.2$, and $S_{1.1\,\rm mm}/S_{1.2\,\rm mm}=1.2$.  Our directly 
calculated counts are in excellent agreement with the rescaled results of 
the AzTEC surveys.  Additionally,  
our prediction for the shape of the number counts below our sensitivity 
threshold, afforded by our \textit{P(D)} analysis, agrees well with the 
deepest AzTEC number counts and is even in rough agreement with the 
deepest SMG counts by \cite{knud08}.

Figure \ref{f-counts} also compares our results to various number count
predictions derived from backward evolution models that incorporate 
multi-waveband observations of number counts and redshift distributions,
as well as limits imposed by the CIB light.  We have restricted this
comparison to models that offer predictions at wavelengths of 1.2\,mm
\citep{beth11} or at 1.1\,mm \citep{vali09,rowa09,mars11},
to which we can apply the rescaling described above.  Although flux
scaling will generally not provide a precise representation of a
model's predictions at 1.2\,mm, the extrapolation from 1.1\,mm to
1.2\,mm is fairly modest.  At flux densities equal to 
or less than those of our significant detections, we find that our 
observations are generally consistent with all model predictions,
although the \citet{vali09} model slightly overpredicts our $P(D)$ curve
near $1\,\rm mJy$.  At the high flux density limit, all models uniformly
overpredict the counts from our best-fitting Schechter function model, 
while remaining consistent with the predictions from our power-law
result (which is only marginally compatible with the CIB; see Figure 
\ref{f-PDspace}).
However, we cannot draw any conclusions from this apparent discrepancy, 
as our $P(D)$ 
analysis cannot constrain the differential counts at flux densities 
greater than those of our brightest detections.

\subsection{Fractional counterpart identification}

Here we investigate the question of whether our radio counterpart 
identification rate ($R_{1.4\,\rm GHz}$) of $\simeq\,93^{+4}_{-7}\%$ $(38/41)$ in the 
LHN is intrinsically greater than is seen in other surveys, or if it is simply 
a function of the increased 20\,cm sensitivity in this field.  We compare our 
identification rate to those found for previous deep surveys at $850\,\mu\rm 
m$, $870\,\rm\mu m$, $1.1\,\rm mm$, and $1.2\,\rm mm$.  Table 
\ref{t-counterparts} lists recent millimeter and submillimeter deep field surveys from the 
GOODS-N, LHE, SXDF, COSMOS, and ECDF-S fields \citep{borys03,borys04,ivis07,bert07,
schin07,perer08,chapi09,weiss09,bigg10}, along with their 20\,cm radio counterpart 
identification rates and 20\,cm map sensitivities.  Because the surveys have 
different definitions of ``significant'' (sub)millimeter 
detections, different data reduction techniques, and different standards for 
radio counterpart associations, we marginalize over all of these variables 
by looking at the average radio counterpart identification rate, and the 
average 20\,cm map sensitivity.  Using the surveys listed in Table 
\ref{t-counterparts}, we find $\langle \sigma_{1.4\,\rm GHz} \rangle 
\simeq\,7.2\, \rm \mu Jy$ and $ \langle R_{1.4\,\rm GHz} \rangle \simeq 
57\%$.

If we imagine that our field had a sensitivity $\langle \sigma_{1.4\,\rm GHz} 
\rangle \simeq\,7.2\,\mu\rm Jy$, six of our likely radio counterparts would 
fall below the $4.0\,\sigma$ limit of $S_{1.4\,\rm GHz} < 29\,\mu \rm Jy$ and 
would not be detected.  Four additional likely 20\,cm counterparts would appear at 
the 4--$5\,\sigma$ level and would be at high risk of not being detected due to the 
usual completeness effects.  Therefore, our radio counterpart identification 
rate would be $68^{+8}_{-9}\%$--$78^{+7}_{-8}\%$.  This range is only marginally 
greater than the average value of 57\%, and well within the scatter of the 
previous surveys.  Therefore, we attribute our high radio identification 
rate to the extremely sensitive VLA map of this field, rather than to 
unusual properties of 1.2\,mm-selected sources at this depth.

Because we expect $\sim 5$ spurious detections among our 41 sources 
with ${\rm S/N} > 4.0$ and we find only 2--3 detections with unlikely or
no radio counterparts, there is little room left to accommodate a 
substantial, extremely high-redshift ($z > 5$) population of radio-undetected 
SMGs \citep[see also][]{ivis05}.  This work suggests that with a deep enough 
radio image, perhaps all SMGs might have their radio counterparts identified, 
auguring well for upcoming deep surveys that exploit the dramatically expanded 
correlator bandwidth of the EVLA.

We find that $7.3^{+6.7}_{-4.0}\%$ (3/41) of our detections have 
two likely radio counterparts (MAMBO ID\,\# 3, 15, and 39).  If we consider 
the fact that $5\%$ of all randomly chosen positions within our MAMBO map 
will have counterparts within $8''$ with $P\leq 0.05$, then we would 
expect to find a double radio counterpart rate of $\sim 4.6\%$ from
chance associations.  
Previous studies have found that $\sim 10\%$ of SMGs host multiple 
likely radio counterparts \citep[see, e.g.,][]{ivis02,ivis07,pope06}, probably
due to the effects of confusion within the submm/mm image, physical 
interactions, or the extended jets of radio-loud AGN.  Although our SMG 
sample in the LHN is too small to be able to constrain the fraction of 
multiple radio counterparts to better than $\pm 5\%$, we note that the 
pair separations of the radio counterparts are $2.1''$, $7.7''$, and $7.4''$,
and that two of the three MAMBO sources have deboosted flux densities in the 
top $25\%$ of our sample.  These results may be in agreement with the trend 
identified in \cite{ivis07} that multiple radio counterparts are preferentially
associated with the brightest SMGs, and have pair angular  separations 
$\Delta\theta\simeq 2''$--$6''$.

\subsection{Redshift distribution} \label{ss-redshift}

\begin{figure}
\epsscale{1.}
\plotone{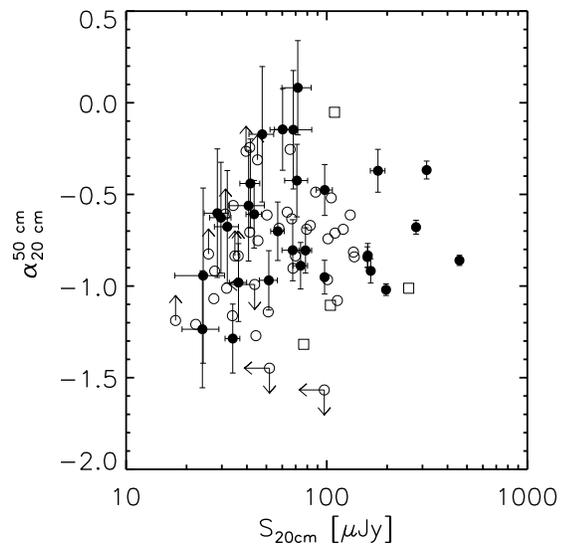}
\caption{50/20\,cm spectral index vs. 20\,cm flux density for SMG radio counterparts.  Solid circles represent the
20 unblended 50\,cm radio counterparts to our MAMBO detections.  Open circles (squares) 
represent non AGN-dominated (AGN-dominated) SMGs in the LHE field \citep{ibar10}.}
\label{f-rindex}
\end{figure}

As listed in Table \ref{t-sources} and detailed in Appendix \ref{a-sources}, 
of our 41 significant individual detections, two have optical spectroscopic 
redshifts \citep{poll06,owen09a}, two have mid-IR spectroscopic redshifts 
\citep{fiol10}, and two have high-quality photometric redshifts based on 
{\it Herschel} far-IR photometry that we will denote in what follows as 
``$z_{\rm phot}^\prime$'' \citep{magd10}.  For those of the remaining 35 
sources with robust or likely radio counterparts, we generally adopt the 
optical photometric redshifts (denoted $z_{\rm phot}$ in what follows) 
determined by \citet{stra10} for the radio catalog of \citet{owen08}.  
The exception to this rule comes for $\{z_{\rm phot}\}$ to which 
\citet{stra10} assign a goodness-of-fit quality flag of ``C''; these 
redshifts are less reliable, and in particular are more likely to manifest 
catastrophic errors.  For such sources, as well as for the one 1.2\,mm 
detection that 
lacks a radio counterpart altogether, we instead derive our own redshift 
estimates ($z_\alpha$) using the radio-submillimeter spectral index redshift 
indicator of \citet{cari99}:
\begin{equation} \label{e-cy99}
\alpha^{350}_{1.4} = -0.24 -  \left[ 0.42 \times (\alpha_{\rm radio} - 
\alpha_{\rm submm}) \times \log_{10}(1+z_\alpha)  \right] 
\end{equation}
\citep[see also][]{cari00,yun02}.
For $\alpha_{\rm radio}$ we use, in order of priority and availability, 
$\alpha^{\rm 90cm}_{\rm 20cm}$, $\alpha^{\rm 50cm}_{\rm 20cm}$, or 
$-0.68$.  We use  $\alpha^{50\,\rm cm}_{20\,\rm cm}$ only for sources with
clean, unblended $50\,\rm cm$ detections.  For these unblended
50\,cm counterparts, we find an average value of 
$\left \langle \alpha^{50\,\rm cm}_{20\,\rm cm} \right \rangle = -0.68\pm 0.06$ 
(see Figure \ref{f-rindex}), in agreement with the average spectral 
index of SMGs in the LHE field of 
$\left \langle \alpha^{50\,\rm cm}_{20\,\rm cm} \right \rangle 
= -0.75\pm 0.06$  \citep{ibar09,ibar10}.  We adopt this mean value 
of $\alpha^{50\,\rm cm}_{20\,\rm cm}$ ($-0.68$) for redshift determination 
of sources with \textit{only} 20\,cm radio counterparts, or whose 
$50\,\rm cm$ counterparts are confused.  For $\alpha_{\rm submm}$, we 
use the spectral index at 1.2\,mm of the fiducial high-redshift dusty 
galaxy SED ($\alpha_{\rm submm}=3.2$), as motivated in \S 
\ref{ss-compare counts}.  For our detection with no likely radio counterpart, 
we estimate a redshift lower bound by using the local 
$S_{\rm 20\,cm}$ $4\sigma$ upper limit in Equation \ref{e-cy99}.

\begin{figure} 
\epsscale{1.}
\plotone{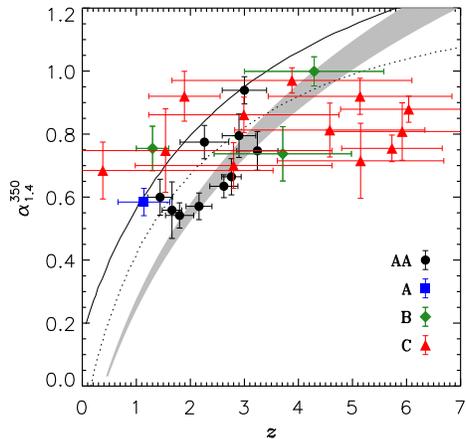}
\caption{350/1.4\,GHz spectral index ($\alpha^{350}_{1.4}$) vs. \textit{z}.  
Black circles, blue squares, green diamonds, and red triangles have $z_{\rm phot}$ with fit 
qualities of AA, A, B, and C, respectively.  
The shaded area shows the expected behavior of $\alpha^{350}_{1.4}$ for the 
\citet{cari99} spectral index redshift indicator, with 
$-\alpha_{\rm radio}$ ranging from 0.52--0.80.  The solid and dashed lines 
show the empirical relations obtained by redshifting the SEDs of nearby 
starburst galaxies Arp\,220 and M82, respectively.}
\label{f-zindex}
\end{figure}

Figure \ref{f-zindex} illustrates why we exclude C-quality 
photometric redshifts from our catalog.  Plotted is the 
$\alpha^{350}_{1.4}$ spectral index of the detections as a function 
of $z_{\rm phot}$.  The points are coded according 
to photometric redshift quality flag \citep{stra10}.  The points with the 
best photometric redshift fit quality (AA) are shown as black circles, followed 
by blue squares (A), green diamonds (B), and red triangles 
(C).  The shaded region shows the \citet{cari99} relation 
for $-\alpha_{\rm radio}=0.52-0.80$, where $-0.52$ represents the median
$\alpha^{90\rm\,cm}_{20\rm\,cm}$ spectral index in the LHN field \citep{owen09b}
and $-0.80$ is the fiducial synchrotron value \citep{cond92}.  The over-plotted 
lines show the empirical relations recovered by redshifting the SEDs of 
nearby star-forming galaxies M82 and Arp\,220 \citep{klei88,scov91}.  
The highest quality photometric redshifts agree with their galaxies' 
spectral indices in that they either follow the Carilli \& Yun relation, 
or are consistent with an M82 or Arp\,220 SED.  In contrast, the 
C-quality photometric redshifts are scattered almost uniformly in $z$
for a given $\alpha^{350}_{1.4}$, demonstrating their lack of robustness.

\begin{figure} 
\epsscale{0.8}
\plotone{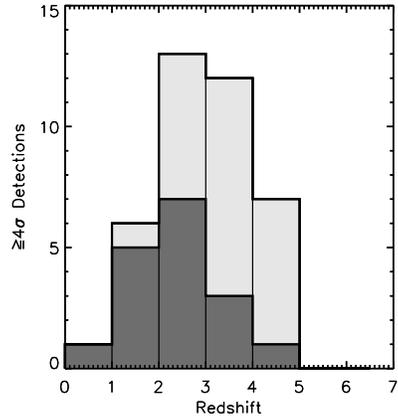}
\caption{Redshift distribution of significant detections.  The light grey 
histogram includes all sources.  The dark grey histogram includes only 
sources with spectroscopic or high-quality ({\it Herschel} or AA/A/B-grade 
optical) photometric redshifts.}
\label{f-redshito}
\end{figure}

Figure \ref{f-redshito} shows the redshift distribution of 
our catalog, including all spectroscopic, photometric, and 
$\alpha^{350}_{1.4}$-estimated redshifts.  It is apparent in Figure 
\ref{f-redshito} that the 17 spectroscopic and high-quality ({\it Herschel} and 
AA/A/B-grade optical) photometric redshifts are biased towards lower 
redshifts.  The median redshift for this $41\%$ of our sample is 
$z_{\rm median} = 
2.26$, with an inter-quartile range of 1.72--2.90. For all galaxies, 
$z_{\rm median}= 2.90$, with an inter-quartile range of 2.33--3.70.  This 
systematic bias has two causes.  First, the highest-redshift galaxies 
have the faintest counterparts, and will necessarily be detected 
in fewer optical bands, which results in a poorer fit.  This trend is in
contrast to the full radio catalog, for which the median $z_{\rm phot}$ is $\sim 1$
and the fraction of photometric redshifts with AA/A/B quality ($\sim85\%$) 
is much higher
than for our MAMBO sources.  Second, the SEDs in the 
\citet{stra10} galaxy template library are most representative of nearby 
galaxies, potentially resulting in a poor fit if they are applied to 
high-\textit{z} galaxies whose SEDs are not included in that library.

The median redshift for our sample ($z_{\rm median}=2.90$) is 
larger than the 
median redshift determined by \citet{pope06} for a complete sample of 
$850\,{\rm \mu m}$-selected SMGs with spectroscopic 
redshifts ($z_{\rm median}=2.0$).  Although our 
redshift distribution has greater uncertainties because it relies heavily 
on photometric redshifts and spectral index redshift estimates, it is
in agreement with the results of \cite{chapi09}, who have shown that, with high
statistical significance, galaxies in a sample selected at 1.1\,mm are 
detected at higher redshift ($z_{\rm median}=2.7$) than those selected at 
$850\,\mu \rm m$.  Our median redshift is also greater than that of the
sample of 68 galaxies selected at $870\,\mu\rm m$ from the
LABOCA survey of the ECDF-S.  Using 17-band optical through mid-IR 
photometry, \cite{ward11} find $z_{\rm median}=2.2$.

\subsection{Spatial correlation with 20\,cm sources} \label{ss-radiocorr}

The results from the $w(\theta)$ and nearest neighbor analyses (\S 
\ref{ss-clustering}) suggest that our sources are clustered to some degree. 
The spike in $w(\theta)$ on $\sim 4'$ scales is an intriguing result that is 
consistent with the visual impression of Figure \ref{f-map} (i.e., the 
southeastern overdensity and the central void) and hints at the existence 
of large-scale structure (LSS) in this field.  To investigate whether the 
spatial distribution of our detections traces LSS that can also 
be seen at other wavelengths, we compare our source positions to the 
distribution of radio sources within the LHN.  In order to compare
our MAMBO sources to radio sources at comparable redshifts, we include only
radio sources with $1.5 < z_{\rm phot} < 4.5$ (excluding all that only have 
a C-quality $z_{\rm phot}$).  Additionally, we only consider radio sources
with sizes greater than $1.0''$.  These larger sources will be preferentially 
gas-rich mergers with extended star formation or radio-loud AGN, which we 
would naively expect to trace environmental overdensities on the basis 
of studies at lower redshift \citep[e.g.,][]{hill91,best04,best05,wake08}.
From a practical standpoint, they can also be detected over the full area 
of the MAMBO map, allowing for a fair comparison; sources with 20\,cm sizes 
$\leq 1''$, in contrast, tend to be fainter, and therefore have systematically 
lower surface densities farther from the center of the VLA map.  

\begin{figure} 
\epsscale{1.}
\plotone{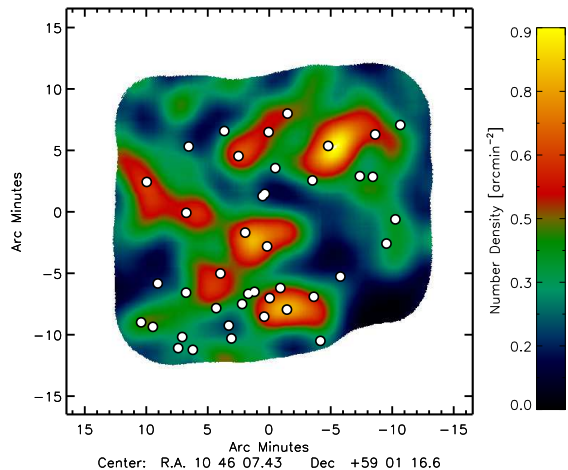}
\caption{Distributions of individually detected MAMBO sources and the 
smoothed surface density of 20\,cm radio sources with $1.5 < z_{\rm phot} < 
4.5$ (excluding radio sources with C-quality photometric redshifts) and sizes 
$\geq 1''$.  The area shown is the region in the ``full'' map used for source
extraction. }
\label{f-radiotrace}
\end{figure}

In Figure \ref{f-radiotrace}, we plot our 1.2\,mm source positions over a 
$3'$-resolution smoothed surface density map of the 307 radio sources in the 
\cite{owen08} VLA catalog that satisfy our selection cuts.  We find that the 
distributions of the two populations agree quite well: (1) the density map 
shows a deficiency of radio sources at the location of our central MAMBO void, 
(2) every radio source density peak ($\langle \Sigma_{\rm 20\,cm} \rangle \geq 
0.5\,{\rm arcmin}^{-2}$) is associated with at least one MAMBO detection, and 
(3) $\sim 66\%$ of our MAMBO detections are located in regions with higher 
than average ($\langle \Sigma_{20\,\rm cm} \rangle \geq 0.34\,\rm 
arcmin^{-2}$) radio source density, whose area comprises only $45\%$ of the 
total.  
This striking agreement seems to argue for a real physical
correlation between our MAMBO detections and $20\,\rm cm$ radio galaxies at 
similar redshifts.  Following \citet{aust09}, we have also compared our
SMG catalog to an identical number of homogeneously distributed, randomly
chosen positions.  The experiment confirms the spatial correlation at a
confidence level of 90\%.  However, when using random positions derived 
from our simulated maps that take into account the spatially varying 
sensitivity, we find 
that our MAMBO detections, although significantly correlated with each 
other, are not significantly spatially correlated with this sample of 
high-\textit{z} $20\,\rm cm$ galaxies.

\subsection{Resolving the {1.2\,\rm {\bf mm}} CIB} \label{ss-CIB}

At 1.2\,mm, the cosmic infrared background (CIB) intensity is $I_{\nu} \simeq 
15-24\,\rm Jy\,deg^{-2}$ \citep{puget96,fixsen98}.  By adding up the deboosted 
flux densities of our detections with ${\rm S/N} >4.0$, we recover $\simeq \rm 
0.58\,Jy\,deg^{-2}$ of the CIB, or about $\sim3\%$.  Figure \ref{f-PDspace} 
shows that our best fitting Schechter function estimate of the differential
number counts is entirely consistent with the intensity of the CIB, while
the power-law model is only marginally compatible with it.  The analysis performed 
by \cite{scot10} on ASTE/AzTEC data in the GOODS-S field finds that 
the best-fit power law model from their \textit{P(D)} analysis can account for 
the CIB, although only if they integrate the counts past the cutoff used 
for that analysis.  They also find that their Schechter function 
model is incompatible with the CIB, and can only recover $\sim 30\%$ of the
$1.1\,\rm mm$ background when integrated down to $S_{1.1\,\rm mm}'=0\,\rm mJy$.
These results may be due to their a priori choice of a faint-end power-law 
index $\delta'=1.0$.  We fit for this parameter directly and find that a 
steeper value of $\delta' \simeq 1.86$ is optimal for our data and produces 
enough faint sources to account fully for the CIB light at 1.2\,mm.

\section{Conclusions}

We have presented a $566\,{\rm arcmin^{2}}$ map of the Lockman Hole North field with 
an average optimally filtered point source sensitivity $\simeq 0.75\,{\rm 
mJy\,beam^{-1}}$.  
By making use of previously developed and original 
techniques to handle chopped bolometer array data, along with 
\textit{P(D)}-based number counts and clustering analyses, we have
assembled a comprehensive picture of the $1.2\,\rm mm$ sky.
Our results provide valuable new constraints for models of the 
evolution of dusty starburst galaxies through cosmic time.

We detect 41 $1.2\,\rm mm$ sources at ${\rm S/N} > 4.0$ 
in our final map.  Of these 41 detections, 38 have robust or likely 
($P < 0.05$) 20\,cm radio counterparts, and 20 have robust or likely 
counterparts at $\rm 24 \, \mu m$.  Based on Monte-Carlo simulations, we 
expect $\sim 5$ of these detections to be spurious, and only $\sim 2$ 
20\,cm counterparts with $P<0.05$ to be chance associations.  This result gives our 
MAMBO/LHN map the highest single-field SMG radio counterpart 
identification rate ever observed ($93^{+4}_{-7}\%$), which we have shown 
can be explained entirely by the extraordinary depth of our 20\,cm VLA map.  
The enhanced sensitivity of the EVLA will be able to make high counterpart
fractions routine for future SMG samples.  Based on the spectroscopic, 
photometric, and radio/far-infrared spectral index redshifts of these 
counterparts, the median 
redshift of our sample is $z_{\rm median}=2.9$, higher than 
has been determined 
for $850\,\rm\mu m$-selected SMG samples in fields with shallower VLA coverage.

We estimate the number counts of $\rm 1.2 \, mm$ sources both directly and 
by using a \textit{P(D)} analysis and find a similar slope but a lower 
overall normalization relative to previous MAMBO surveys.  However, our 
results are in close agreement, after a scaling in flux density, with 
those of recent surveys at $1.1\,\rm mm$.  The compatibility of  
our directly calculated counts and \textit{P(D)} analysis with the constraint 
of the $1.2\,\rm mm$ CIB demonstrate the robustness of our 
results.  In particular, we find that for 
$S_{1.2\,\rm mm}\lesssim 0.05\,\rm mJy$ the SMG differential number 
counts cannot keep rising with the faint-end slope observed for $S_{1.2\,\rm mm}> 
0.05\,\rm mJy$, and that the bright SMG population contributes at most a small 
fraction to the $1.2\,\rm mm$ CIB.  

The high resolution afforded by the IRAM 30\,m telescope, the large extent
of our map, and the use of analysis  methods that thoroughly take into 
account the negative residuals of the chopped triple-beam PSF have allowed 
us to demonstrate possible clustering in 
the  $1.2\,\rm mm$ population.  The SMG correlation function, a nearest neighbors 
analysis, and, to a lesser extent, the spatial correlation of 
our significant detections 
with large radio sources over the same redshift range all suggest that 
our sample traces some degree of large scale structure at high redshift.
Our work prepares the $1.2\,\rm mm$ waveband for the ALMA era by 
creating a better understanding of this population's statistical properties
and setting new $1.2\,\rm mm$ constraints for galaxy evolution models.

\acknowledgments

We thank the scientific and technical staff at IRAM, particularly Robert 
Zylka, as well as several seasons' pool observers, for their help in making 
this project a success.  We thank Jacqueline Bergeron, 
David Hughes, Chuck Keeton, Maurilio Pannella, and Jean Walker for useful 
discussions.  We also thank the referee for very useful comments that 
have improved the analysis and results of this paper.  This project has been 
supported by NSF grant AST-0708653.

\medskip

{\it Facilities:} \facility{IRAM}, \facility{VLA}, \facility{SST}, \facility{GMRT}.

\appendix

\section{Notes on individual detections} \label{a-sources}

\subsection{MM\,J104700.1+590109 = ID \# 1}

\citet{poll06} report an optical $z_{\rm spec} = 2.562$ for this source.
It also has a $70\,\mu\rm m$ counterpart with 
$S_{70\,\mu\rm m}=10.4\pm1.7\,\rm mJy$ at a distance of $3.2''$, and a
$160\,\mu\rm m$ counterpart with $S_{160\,\mu\rm m}=24.1\pm
1.9\,\rm mJy$ at a distance of $1.7''$ \citep{owen11b}.

\subsection{MM\,J104631.4+585056 = ID \# 3}

\begin{figure} 
\epsscale{1.}
\plotone{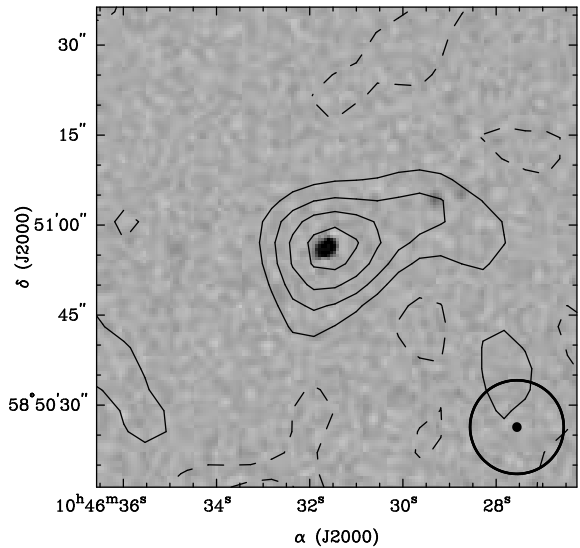}
\caption{1.2\,mm contours (multiples of $1.0\,{\rm mJy\,beam^{-1}}$) for 
MAMBO source MM\,J104631.4+585056 = ID \# 03, overlaid on 20\,cm map showing
double counterpart.  The black 
circle and small filled ellipse at lower right represent the $15.6''$ MAMBO 
beam (after filtering) and the $1.63'' \times 1.57''$ VLA beam, respectively.}
\label{f-source3}
\end{figure}

In addition to the robust 20\,cm counterpart listed in Table \ref{t-sources}, 
this source has an additional likely counterpart with $S_{\rm 20\,cm} = 
30\,{\rm \mu Jy}$, separation $2.6''$, and $P = 0.034$, with which it is 
nearly blended.  Neither radio counterpart has an estimated photometric 
redshift.  Figure \ref{f-source3} shows 1.2\,mm contours overlaid on a 
20\,cm cutout image that includes both counterparts.

\subsection{MM\,J104638.4+585613 = ID \# 6}

We identify this source with LHN8 in the {\it Herschel} catalog of 
\citet{magd10}, from which it is separated by $2.0''$ ($P = 0.0084$), and with 
SWIRE4\_J104638.68+585612.5 = ID \# L14 in the {\it Spitzer} sample of 
\citet{fiol10}, who report a mid-IR $z_{\rm spec} = 2.03$.  
\citet{fiol09} report an on-off flux density measurement of $S_{\rm 1.2\,{\rm 
mm}} = 2.13 \pm 0.71\,{\rm mJy}$, which is consistent with our (non-deboosted) 
$S_{\rm 1.2\,mm} = 2.7 \pm 0.5\,{\rm mJy}$ within the uncertainties.

\subsection{MM\,J104704.9+585008 = ID \# 9}

The radio counterpart to this source ($P=0.0043$) is not in the catalog of 
\citet{owen08} because it has a S/N ratio of 4.8.

\subsection{MM\,J104556.5+585317 = ID \# 11}

We identify this source with LHN1 in the {\it Herschel} catalog of 
\citet{magd10}, from which it is separated by $3.4''$ ($P = 0.0057$), and with 
SWIRE4\_J104556.90+585318.8 = ID \# L11 in the {\it Spitzer} sample of 
\citet{fiol10}, who report a mid-IR $z_{\rm spec} = 1.95$ in good 
agreement with the optical $z_{\rm phot} = 1.80$ reported by \citet{stra10}.
\citet{fiol09} report an on-off flux density measurement of $S_{\rm 1.2\,{\rm 
mm}} = 3.08 \pm 0.58\,{\rm mJy}$, which is consistent with our (non-deboosted) 
$S_{\rm 1.2\,mm} = 3.4 \pm 0.6\,{\rm mJy}$ within the uncertainties.  This source
also has a $160\,\mu\rm m$ counterpart \citep{owen11b} with $S_{160\,\mu\rm m}=
11.8\pm1.5\,\rm mJy$ at a distance of $4.3''$

\subsection{MM\,J104728.3+585213 = ID \# 15}

\begin{figure} 
\epsscale{1.}
\plotone{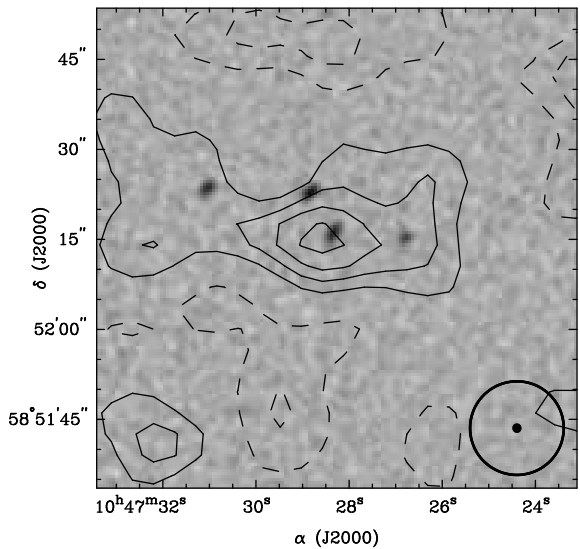}
\caption{1.2\,mm contours (multiples of $1.0\,{\rm mJy\,beam^{-1}}$) for 
MAMBO source MM\,J104728.3+585213 = ID \# 15, overlaid on 20\,cm map showing
multiple counterparts.  Other notation is as in Figure \ref{f-source3}.}
\label{f-source15}
\end{figure}

This source has two radio counterparts; the primary counterpart listed in 
Table \ref{t-sources} has $z_{\rm phot} = 2.76$, while the second (with $P = 
0.032$) has $z_{\rm phot} = 1.06$.  In the 20\,cm map, we see a quadruple 
radio source (see Figure \ref{f-source15}).  At a low level of significance,
the 1.2\,mm emission appears to be elongated in the same direction as the 
radio source(s).  This system also has a $160\,\mu\rm m$ counterpart 
\citep{owen11b} with $S_{160\,\mu\rm m}=12.7\pm1.5\,\rm mJy$ at a distance of $6.8''$.

\subsection{MM\,J104610.4+590242 = ID \# 17}

This source is separated by only $\sim 15''$ from MM\,J104611.9+590231 = ID \# 
39.  The catalog of \citet{owen08} includes a single radio source whose 
nominal position is midway between two very faint sources (one resolved, one 
unresolved) that are visible in the original 20\,cm map (see Figure 
\ref{f-source17_39}).  These two sources, with peak flux densities of 
$15.6\,\rm \mu Jy$ and $13.5\,\rm \mu Jy$, are only identifiable because they 
lie very close to the center of the VLA map, where the local RMS noise is only 
$2.9\,\rm \mu Jy\,beam^{-1}$.  After attributing the flux of the catalogued 
source to its two constituents, we find that one is a robust radio counterpart 
for the MAMBO source ($1.0''$ separation with $P=0.0099$), while the other is 
probably a chance association ($3.3''$ separation with $P=0.077$).

\subsection{MM\,J104617.0+585444 = ID \# 20}

This source has an unlikely 20\,cm radio counterpart ($P=0.12$) with 
$\rm S/N=4.2$ ($S_{\rm 20\,cm}=15.0\pm3.6\,\rm\mu Jy$), and is therefore 
not in the catalog of \cite{owen08}, which includes only sources 
with $\rm S/N>5.0$.

\subsection{MM\,J104522.8+585558 = ID \# 26}

This source has no radio counterpart within $8''$ and no likely $24\mu \rm m$ 
counterpart, although it does have an X-ray counterpart (see \S \ref{ss-xray}).

\subsection{MM\,J104620.9+585434 = ID \# 28}

This source has a likely 20\,cm radio counterpart ($P=0.029$) with 
$\rm S/N=4.3$ ($S_{\rm 20\,cm}=24.8\pm5.8\,\rm\mu Jy$), and is therefore not 
in the catalog of \cite{owen08}, which includes only sources with $\rm S/N>5.0$.

\subsection{MM\,J104556.1+590914 = ID \# 29}
\label{id-29}
\begin{figure} 
\epsscale{1.}
\plotone{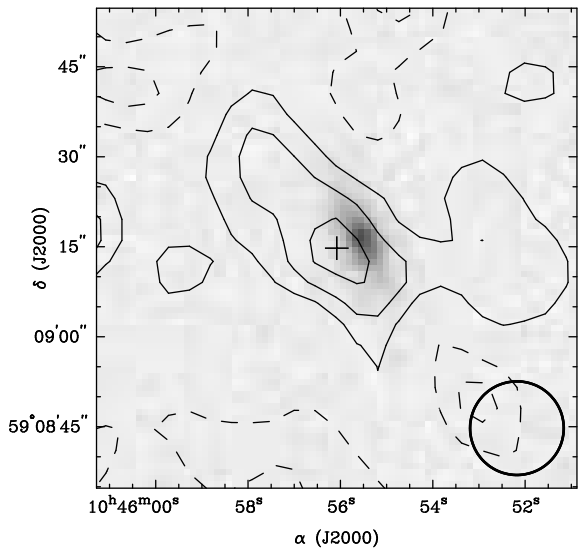}
\caption{1.2\,mm contours (multiples of $0.8\,{\rm mJy\,beam^{-1}}$) for 
MAMBO source MM\,J104556.1+590914 = ID \# 29, overlaid on a DSS red image 
of SDSS\,J104555.49+590915.9 at $z_{\rm spec} = 0.044$ \citep{owen09a}.  
Other notation is as in Figure \ref{f-source3}.}
\label{f-source29}
\end{figure}

We identify this source with SDSS\,J104555.49+590915.9, an optically bright 
galaxy for which \citet{owen09a} report an optical $z_{\rm spec} = 0.044$.  
Figure \ref{f-source29} shows a red optical image overlaid with 1.2\,mm 
contours, which at a low level of significance are elongated in the same 
direction as the galaxy's stars.  This $\simeq 20\,''$ source is heavily 
resolved at $20\,\rm cm$ and $50\,\rm cm$.  It is also detected at 
$160\,\mu\rm m$ \citep{owen11b} with $S_{160\,\mu\rm m}=
15.8\pm1.4\,\rm mJy$ (at a separation of $6.6''$), and at $250\,\rm\mu m$
 \citep[\textit{Herschel}/SPIRE;][]{smit10} with 
$S_{250\rm\,\mu m}=133\pm8\,\rm mJy$ (at a separation of $3.6''$).

\subsection{MM\,J104539.6+585419 = ID \# 32}

We identify this source with LHN3 in the {\it Herschel} catalog of 
\citet{magd10}, from which it is separated by $5.2''$ ($P = 0.016$).  
\citet{magd10} estimate $z_{\rm phot}^\prime = 2.40$ on the basis of their 
PACS and SPIRE photometry, which we list in Table \ref{t-sources} rather than 
the \citet{stra10} optical $z_{\rm phot} = 1.32$, due to the close connection 
between far-IR and millimeter emission.

\subsection{MM\,J104608.1+590744 = ID \# 36}

This source has a 20\,cm radio counterpart ($P=0.041$) with 
$\rm S/N=4.9$ ($S_{\rm 20\,cm}=16.1\pm3.3\,\rm\mu Jy$), and 
is therefore not in the catalog of \cite{owen08}, which includes only sources 
with $\rm S/N>5.0$.

\subsection{MM\,J104610.8+585242 = ID \# 37}

We identify this source with LHN4 in the {\it Herschel} catalog of 
\citet{magd10}, from which it is separated by $2.4''$ ($P = 0.013$).  
\citet{magd10} estimate $z_{\rm phot}^\prime = 1.72$ on the basis of their 
PACS and SPIRE photometry, which we list in Table \ref{t-sources}; this is in 
good agreement with the optical $z_{\rm phot} = 1.66$ reported by 
\citet{stra10}.

\subsection{MM\,J104611.9+590231 = ID \# 39}

\begin{figure} 
\epsscale{1.}
\plotone{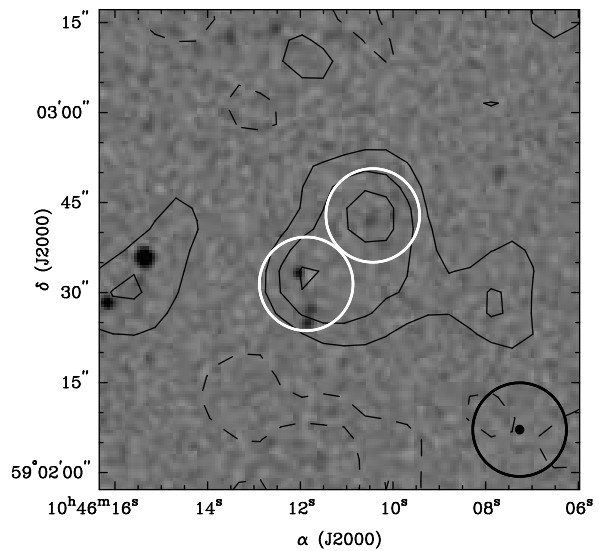}
\caption{Nearly blended MAMBO sources MM\,J104610.4+590242 (ID \# 17), which 
has one likely 20\,cm counterpart, and MM\,J104611.9+590231 (ID \# 39), which 
has two likely 20\,cm counterparts.  1.2\,mm contours (multiples of 
$0.7\,{\rm mJy\,beam^{-1}}$) are overlaid on 20\,cm greyscale.  White circles 
are centered on the positions of the extracted MAMBO sources; other notation 
is as in Figure \ref{f-source3}.}
\label{f-source17_39}
\end{figure}

This source is separated by only $\sim 15''$ from MM\,J104610.4+590242 = ID 
\# 17.  It also has a pair of likely radio counterparts within an 
$8''$ search radius.  Figure \ref{f-source17_39} shows 1.2\,mm contours 
overlaid on a 20\,cm cutout image.  This source is not identified in the
catalog of \cite{owen08}.


\clearpage

\clearpage

\clearpage

\begin{deluxetable}{lrrccc}
\tablecaption{MAMBO observations}
\tablewidth{0pt}

\tablehead{
\colhead{ } & \colhead{ } & \colhead{ } & \colhead{Tracking} 
& \colhead{Bad refraction} & \colhead{Chop throws} \\
\colhead{Season} & \colhead{Maps} & \colhead{Hours} & \colhead{jitter?} 
& \colhead{correction?} & \colhead{(arcsec)}}
\startdata
2006 winter    &  76 &  78.9 & X & X & 36/48 \\
2008 summer    &  12 &  12.3 & X &   & 42/36 \\
\hline
2008 winter    &  39 &  40.3 &   &   & 42/36 \\
2009 summer    &   8 &   8.0 &   &   & 42/36 \\
2009 winter    &  52 &  53.0 &   &   & 42/36 \\
\hline
Total ``best'' &  99 & 101.3 &   &   & \\
Total ``full'' & 183 & 192.5 &   &   & \\
\enddata
\tablecomments{``Best'' data use maps from winter 2008, summer 2009, and 
winter 2010.  ``Full'' data use maps from all seasons in the table.}
\label{t-obs}
\end{deluxetable}

\clearpage

\begin{deluxetable}{cccccccccccccc}

\tabletypesize{\scriptsize}

\rotate

\tablecaption{MAMBO detections}
\tablewidth{0pt}
\setlength{\tabcolsep}{0.03in}

\tablehead{ 
  \colhead{} & 
  \colhead{} &
  \colhead{$S_\nu^{\rm Best\tablenotemark{a}}$} & 
  \colhead{$S_\nu^{\rm Full\tablenotemark{b}}$} & 
  \colhead{$S_\nu^{\mbox{\begin{tiny}Deboosted\end{tiny}}\tablenotemark{c}}$ } &
  \colhead{} & 
  \colhead{} & 
  \colhead{$D_{\rm 20\,cm}^{\,\tablenotemark{e}}$} & 
  \colhead{} & 
  \colhead{$S_{\rm 20\,cm}^{\,\tablenotemark{g}}$} &
  \colhead{$S_{\rm 50\,cm}^{\,\tablenotemark{h}}$} &
  \colhead{$S_{\rm 90\,cm}^{\,\tablenotemark{i}}$} &
  \colhead{} & 
  \colhead{$S_{\rm 24\,\mu m}^{\,\tablenotemark{k}}$} 
  \\ 
  \colhead{ID} &
  \colhead{Source name} & 
  \colhead{(mJy)} & 
  \colhead{(mJy)} & 
  \colhead{(mJy)} & 
  \colhead{$P(<0)^{\,\tablenotemark{d}}$} & 
  \colhead{$z$} & 
  \colhead{(arcsec)} &  
  \colhead{$P_{\rm 20\,cm}^{\,\tablenotemark{f}}$} & 
  \colhead{($\rm \mu Jy$)} & 
  \colhead{($\rm \mu Jy$)} & 
  \colhead{($\rm \mu Jy$)} & 
  \colhead{$P_{\rm 24\,\mu m}^{\,\tablenotemark{j}}$} &
  \colhead{($\rm \mu Jy$)} 
   
}

\startdata
  &                    &             &             &                    & \multicolumn{3}{l}{$\rm S/N \ge 4.5$}  \\ 
\tableline

 1&MM\,J104700.1+590109& 3.7$\pm$0.8 &  4.1$\pm$0.6& 3.5$^{+0.6}_{-0.6}$& $<0.01$ & 2.562\tablenotemark{l} & 2.0 & {\bf 0.0025     }    & 278$\pm$6 & $490\pm11$     &  687$\pm$72 & {\bf 0.0018}  & $1280\pm9$  \\ 
 2&MM\,J104627.1+590546& 4.5$\pm$0.8 &  4.7$\pm$0.7& 3.8$^{+0.7}_{-0.7}$& $<0.01$ & 4.29\tablenotemark{m} & 3.0 & {\bf 0.037      }   &  35$\pm$6 & $49\pm 23$\tablenotemark{t}  &             & 0.11         & $41\pm6$    \\ 
 3&MM\,J104631.4+585056& 6.1$\pm$1.8 &  4.7$\pm$0.7& 3.8$^{+0.8}_{-0.7}$& $<0.01$ & 1.8\tablenotemark{n}  & 0.5 & {\bf 0.0001}\tablenotemark{r}   & 458$\pm$9 & $937\pm10$     & 1633$\pm$85 &              &             \\ 
 4&MM\,J104607.4+585413& 2.8$\pm$0.8 &  3.2$\pm$0.5& 2.7$^{+0.5}_{-0.5}$& $<0.01$ & 4.4\tablenotemark{n}  & 0.5 & {\bf 0.0025     }   &  29$\pm$3 & $50\pm11$      &             &               &             \\ 
 5&MM\,J104725.2+590339& 4.9$\pm$0.9 &  5.2$\pm$0.8& 4.0$^{+0.8}_{-0.9}$& $<0.01$ & 3.00\tablenotemark{m} & 3.4 & {\bf 0.028      }   &  51$\pm$5 & $115\pm10$     &             & {\bf 0.0099}  & $395\pm26$  \\ 
 6&MM\,J104638.4+585613& 3.1$\pm$0.7 &  2.7$\pm$0.5& 2.3$^{+0.4}_{-0.4}$& $<0.01$& 2.03\tablenotemark{o} & 2.1 & {\bf 0.0043     }    & 159$\pm$5 & $321\pm9$      &  442$\pm$70 & {\bf 0.0074}  & $662\pm8$   \\ 
 7&MM\,J104700.1+585439& 2.8$\pm$0.7 &  2.8$\pm$0.5& 2.3$^{+0.4}_{-0.5}$& $<0.01$ & 4.2\tablenotemark{n}  & 2.0 & {\bf 0.016      }   &  41$\pm$4 & $60\pm10$      &             & 0.10          & $329\pm18$  \\ 
 8&MM\,J104633.1+585159& 4.5$\pm$1.2 &  3.4$\pm$0.6& 2.7$^{+0.6}_{-0.6}$& $<0.01$ & 3.3\tablenotemark{n}  & 0.4 & {\bf 0.0003     }   &  97$\pm$9 & $145\pm10$     &             & {\bf 0.0022}  & $342\pm8$   \\ 
 9&MM\,J104704.9+585008& 5.6$\pm$1.5 &  5.1$\pm$0.9& 3.8$^{+1.0}_{-0.9}$& $<0.01$ & 3.9\tablenotemark{n}  & 0.6 & {\bf 0.0043}\tablenotemark{s}   &  23$\pm$5 & $67\pm 11$     &             &               &             \\ 
10&MM\,J104622.9+585933& 3.6$\pm$0.7 &  2.9$\pm$0.5& 2.4$^{+0.5}_{-0.5}$& $<0.01$ & 2.6\tablenotemark{n}  & 2.1 & {\bf 0.0083     }   &  78$\pm$5 & $153\pm11$     &  383$\pm$73 & 0.14          & $221\pm8$   \\ 
11&MM\,J104556.5+585317& 3.5$\pm$0.9 &  3.4$\pm$0.6& 2.7$^{+0.6}_{-0.6}$& $<0.01$ & 1.95\tablenotemark{o} & 3.5 & {\bf 0.0057     }   & 314$\pm$10& $427\pm9$      &  662$\pm$74 & {\bf 0.017}   & $684\pm7$   \\ 
12&MM\,J104448.0+590036& 5.1$\pm$0.9 &  3.5$\pm$0.6& 2.7$^{+0.6}_{-0.7}$& $<0.01$ & 2.16\tablenotemark{m} & 3.0 & {\bf 0.0049     }   & 273$\pm$13& $421\pm20$     &  815$\pm$140& {\bf 0.037}   & $597\pm27$  \\ 
13&MM\,J104609.0+585826& 2.7$\pm$0.7 &  2.7$\pm$0.5& 2.1$^{+0.5}_{-0.5}$& $<0.01$ & 1.14\tablenotemark{m} & 3.1 & {\bf 0.0067     }   & 197$\pm$3 & $461\pm10$     & 1101$\pm$151& 0.28          & $128\pm8$   \\ 
14&MM\,J104636.1+590749& 4.3$\pm$0.8 &  4.3$\pm$0.8& 3.0$^{+0.9}_{-0.9}$& $<0.01$ & 2.26\tablenotemark{m} & 2.4 & {\bf 0.0079     }   &  97$\pm$3 & $215\pm14$     &             & {\bf 0.0055}  & $596\pm7$   \\ 
15&MM\,J104728.3+585213& 5.1$\pm$1.1 &  4.5$\pm$0.9& 3.0$^{+0.9}_{-1.0}$& $<0.01$ & 2.76\tablenotemark{m} & 2.4 & {\bf 0.0044}\tablenotemark{r}   & 180$\pm$15& $245\pm12$     &             & {\bf 0.0081}  & $834\pm23$  \\ 
16&MM\,J104720.9+585151& 2.3$\pm$1.0 &  3.8$\pm$0.7& 2.7$^{+0.8}_{-0.8}$& $<0.01$ & 4.9\tablenotemark{n}  & 1.5 & {\bf 0.0088     }    &  47$\pm$6 &  $55\pm15$     &            &  0.13         & $307\pm9$   \\ 
17&MM\,J104610.4+590242& 3.4$\pm$0.7 &  2.6$\pm$0.5& 2.0$^{+0.5}_{-0.5}$& $<0.01$ & 4.0\tablenotemark{n}  & 1.1 & {\bf 0.0099}\tablenotemark{s}    &  27$\pm$3 &  $<45$         &            & {\bf 0.025}   & $293\pm19$  \\ 
18&MM\,J104655.7+585000&             &  4.6$\pm$0.9& 2.9$^{+1.1}_{-1.1}$&$0.017$  & 1.30\tablenotemark{m} & 5.0 & {\bf 0.024      }    & 104$\pm$6 & $202\pm 11$\tablenotemark{t} & 385$\pm$73 & {\bf 0.026}   & $679\pm9$   \\ 
19&MM\,J104502.1+590404&             &  2.6$\pm$0.5& 2.0$^{+0.6}_{-0.6}$& $<0.01$ & 4.1\tablenotemark{n}  & 1.8 & {\bf 0.020      }    &  28$\pm$4 & $47\pm12$      &             &               &             \\ 
20&MM\,J104617.0+585444&             &  2.3$\pm$0.5& 1.7$^{+0.5}_{-0.5}$& $<0.01$ &$>4.6$\tablenotemark{n}  & 5.0 & 0.12\tablenotemark{s} &  $15\pm4$ &  $<33$         &             & {\bf 0.027}  & $67\pm8$    \\ 
21&MM\,J104530.3+590636&             &  2.4$\pm$0.5& 1.8$^{+0.5}_{-0.5}$& $<0.01$ & 3.1\tablenotemark{n}  & 0.9 & {\bf 0.0057     }    &  36$\pm$3 &  $82\pm12$     &             & {\bf 0.017}  & $196\pm8$   \\ 
22&MM\,J104603.8+590448&             &  2.7$\pm$0.6& 2.0$^{+0.6}_{-0.6}$& $<0.01$ & 1.44\tablenotemark{m} & 3.4 & {\bf 0.0094     }    & 165$\pm$5 & $355\pm15$     &  485$\pm$72 & {\bf 0.019}  & $595\pm9$   \\ 
23&MM\,J104641.0+585324&             &  2.2$\pm$0.5& 1.7$^{+0.5}_{-0.5}$& $<0.01$ & 3.6\tablenotemark{n}  & 1.6 & {\bf 0.016      }    &  31$\pm$4 &  $56\pm12$     &             &              &             \\ 
24&MM\,J104500.5+590731&             &  2.4$\pm$0.5& 1.8$^{+0.6}_{-0.5}$& $<0.01$ & 3.24\tablenotemark{m} & 2.3 & {\bf 0.011      }    &  67$\pm$7 & $132\pm10$     &             & 0.089        & $264\pm18$  \\ 
25&MM\,J104540.3+590347&             &  2.3$\pm$0.5& 1.7$^{+0.6}_{-0.6}$& $<0.01$ & 3.5\tablenotemark{n}  & 2.3 & {\bf 0.036      }    &  24$\pm$6 &  $53\pm15$     &             & {\bf 0.0071} & $119\pm7$   \\ 
26&MM\,J104522.8+585558&             &  2.5$\pm$0.6& 1.8$^{+0.7}_{-0.6}$&$<0.01$  &$>5.0$\tablenotemark{n}  & &                            & $<14$     & $<40 $         &             & 0.11         & $239\pm20$  \\ 
27&MM\,J104702.4+585102&             &  3.1$\pm$0.7& 2.0$^{+0.8}_{-0.9}$&$0.018$  & 2.9\tablenotemark{n}  & 0.2 & {\bf 0.0002     }    &  77$\pm$12&  $49\pm12$     &             & {\bf 0.030}  & $237\pm9$   \\ 
\tableline 
  &                    &             &             &                    & \multicolumn{3}{l}{$4.0\le \rm S/N < 4.5$}  \\ 
\tableline
28&MM\,J104620.9+585434&             &  2.1$\pm$0.5& 1.5$^{+0.5}_{-0.5}$& $<0.01$ & 3.8\tablenotemark{n}  & 2.1 &  {\bf 0.029}\tablenotemark{s}    &  $25\pm6$ &     $<60$      &             &               &             \\ 
29&MM\,J104556.1+590914&             &  2.8$\pm$0.7& 1.8$^{+0.8}_{-0.8}$&$0.022$  & 0.044\tablenotemark{p}& 4.7 & {\bf 0.0092     }    & 307$\pm$39& $567\pm 85$  &             & {\bf 0.0012}  & $4838\pm23$ \\ 
30&MM\,J104510.3+590408&             &  2.2$\pm$0.5& 1.5$^{+0.6}_{-0.6}$& $0.011$ & 2.4\tablenotemark{n}  & 1.2 & {\bf 0.0034     }    & 74$\pm$6& $155\pm 10$    &              & {\bf 0.012}   & $56\pm7$    \\ 
31&MM\,J104624.7+585344&             &  2.1$\pm$0.5& 1.5$^{+0.6}_{-0.5}$& $<0.01$ & 2.90\tablenotemark{m} & 2.9 & {\bf 0.027      }    &  43$\pm$3& $72\pm 9$      &              & 0.079         & $236\pm7$   \\ 
32&MM\,J104539.6+585419&             &  2.5$\pm$0.6& 1.7$^{+0.8}_{-0.7}$&$0.023$  & 2.40\tablenotemark{q} & 5.2 &   0.056              &  46$\pm$3& $113\pm9$      &              & 0.059         & $485\pm7$   \\ 
33&MM\,J104535.5+585044&             &  5.2$\pm$1.2& 2.7$^{+1.2}_{-1.2}$&$0.20$   & 3.7\tablenotemark{n}  & 6.3 & {\bf 0.049      }    & 70$\pm$9 & $101\pm 10$    &             & 0.19          & $245\pm7$   \\ 
34&MM\,J104453.7+585838&             &  2.4$\pm$0.6& 1.5$^{+0.7}_{-0.7}$&$0.022$  & 3.5\tablenotemark{n}  & 3.0 & {\bf 0.031      }    &  40$\pm$8 & $65\pm 10$     &             & 0.13          & $76\pm7$    \\ 
35&MM\,J104717.9+585523&             &  2.4$\pm$0.6& 1.5$^{+0.7}_{-0.7}$&$0.025$  & 3.71\tablenotemark{m} & 2.4 & {\bf 0.014      }    & 60$\pm$5 & $68\pm 11$     &              & 0.099         & $297\pm23$  \\ 
36&MM\,J104608.1+590744&             &  2.9$\pm$0.7& 1.7$^{+1.1}_{-0.9}$&$0.052$  & 4.5\tablenotemark{n}   & 2.1 & {\bf 0.041}\tablenotemark{s}    & $16\pm3$ & $<47$          &              & 0.11          & $396\pm39$  \\ 
37&MM\,J104610.8+585242&             &  2.3$\pm$0.6& 1.5$^{+0.8}_{-0.7}$&$0.030$  & 1.72\tablenotemark{q} & 2.3 & {\bf 0.0052     }    & 160$\pm$7& $320\pm10$     &  450$\pm$75 & {\bf 0.026}    & $401\pm8$   \\ 
38&MM\,J104444.5+590817&             &  2.8$\pm$0.7& 1.5$^{+0.9}_{-0.9}$&$0.052$  & 3.6\tablenotemark{n}  & 2.8 & {\bf 0.017      }    &  68$\pm$16& $77\pm10$      &             & 0.12          & $153\pm8$   \\ 
39&MM\,J104611.9+590231&             &  2.1$\pm$0.5& 1.4$^{+0.7}_{-0.6}$&$0.023$  & 2.6\tablenotemark{n}  & 2.2 & {\bf 0.024}\tablenotemark{r}\tablenotemark{s}     &  34$\pm$3 & $99\pm13$       &            & {\bf 0.024}   & $634\pm8$   \\ 
40&MM\,J104658.7+590633&             &  3.0$\pm$0.8& 1.5$^{+1.2}_{-1.1}$&$0.074$  & 2.9\tablenotemark{n}  & 2.6 & {\bf 0.017      }    &  56$\pm$4 & $102\pm11$     &             & {\bf 0.032}   & $229\pm8$  \\ 
41&MM\,J104600.7+585502&             &  2.1$\pm$0.5& 1.4$^{+0.7}_{-0.6}$&$0.025$  & 2.6\tablenotemark{n}  & 3.1 & {\bf 0.018      }    &  71$\pm$11& $67\pm 9$      &             &               &            \\ 

\enddata

\tablecomments{$P$ values in boldface type denote a likely counterpart ($P<0.05$).}

\tablenotetext{a}{Raw flux density extracted from our ``best'' map (see \S \ref{ss-hinge}).}
\tablenotetext{b}{Raw flux density extracted from our ``full'' map(see \S \ref{ss-hinge}).}
\tablenotetext{c}{Flux density extracted from our ``full'' map, and corrected for flux boosting (see \S \ref{ss-boost}).}
\tablenotetext{d}{Total probability that the deboosted flux density is $\le 0\,\rm mJy$ (see \S \ref{ss-boost})}
\tablenotetext{e}{Angular separation between MAMBO detection and 20\,cm counterpart.}
\tablenotetext{f}{Probability of chance association with 20\,cm counterpart.}
\tablenotetext{g}{Flux density of 20\,cm counterpart \citep{owen08}.}
\tablenotetext{h}{Flux density of 50\,cm counterpart \citep{owen11a}.}
\tablenotetext{i}{Flux density of 90\,cm counterpart \citep{owen09b}.}
\tablenotetext{j}{Probability of chance association with $24\,\rm\mu m$ counterpart.}
\tablenotetext{k}{Flux density of $24\,\rm\mu m$ counterpart \citep{owen11b}.}
\tablenotetext{l}{$z_{\rm spec}$ from \cite{poll06}.}
\tablenotetext{m}{$z_{\rm phot}$ from \cite{stra10}.}
\tablenotetext{n}{$z_\alpha$ estimated from \citet{cari99} spectral index (see \S \ref{ss-redshift}).}
\tablenotetext{o}{$z_{\rm spec}$ from \cite{fiol10}.}
\tablenotetext{p}{$z_{\rm spec}$ from \cite{owen09a}.}
\tablenotetext{q}{$z_{\rm phot}'$ from \cite{magd10}.}
\tablenotetext{r}{There are two radio counterparts within $8''$ of MAMBO source with $P<0.05$; see Appendix \ref{a-sources}.}
\tablenotetext{s}{The catalog of \cite{owen08} does not contain this radio counterpart (see \S \ref{ss-20cm}).}
\tablenotetext{t}{Flux density is uncertain due to blending (see \S \ref{ss-50cm}).}

\label{t-sources}
\end{deluxetable}

\begin{deluxetable}{ccccc}
\tablecaption{1.2\,mm number counts} 
\tablewidth{0pt}            
\tablehead{ &  \multicolumn{2}{c}{Differential} & \multicolumn{2}{c}{Integral}  \\
    \colhead{Flux Bin} & \colhead{Flux Density} & \colhead{$dN/dS$}& \colhead{Flux Density} & \colhead{$N(\,>S)$} \\ 
    \colhead{(mJy)} & \colhead{(mJy)}& \colhead{($\rm deg^{-2}\,mJy^{-1}$)} & \colhead{(mJy)}& \colhead{$\rm (deg^{-2})$}}
\startdata

1.68--2.14  & 1.91 & $435^{+369}_{-228}$ & 1.68 & $   366^{+   212}_{-   122}$ \\
2.14--2.59  & 2.36 & $128^{+200}_{-94} $ & 2.14 & $   166^{+   128}_{-    63}$ \\
2.59--3.05  & 2.82 & $156^{+151}_{-89} $ & 2.59 & $    108^{+    90}_{-   46}$ \\
3.05--3.51  & 3.28 & $18^{+80}_{-17}   $ & 3.05 & $    37^{+    58}_{-    22}$ \\
3.51--3.97  & 3.74 & $62^{+97}_{-45}   $ & 3.51 & $    28^{+    44}_{-    21}$
\enddata
\label{t-counts}
\end{deluxetable}

\begin{deluxetable}{llcccc}
\tabletypesize{\scriptsize}
\tablecaption{Area surveyed}
\tablewidth{0pt}
\setlength{\tabcolsep}{0.04in} 
\tablehead{ \colhead{} & \colhead{} & \colhead{} & \colhead{Area} & \colhead{Depth} & \colhead{HPBW} \\
            \colhead{Reference}          & \colhead{Instrument} &           \colhead{Field}      & \colhead{($\rm deg^2$)}   & \colhead{($\rm mJy\,beam^{-1}$)} & \colhead{(arcsec)}
}
\startdata
\cite{grev04}  & MAMBO $1.2\,\rm mm$           & ELAIS-N2 \& LH        &  0.099 & 0.8       & 11   \\
\cite{bert07}  & MAMBO  $1.2\,\rm mm$          & COSMOS             &  0.11  & 1.0       & 11   \\
\cite{scot08}  & AzTEC/JCMT $1.1 \,\rm mm$     & COSMOS             &  0.15  & 1.3       & 18   \\
\cite{perer08} & AzTEC/JCMT $1.1 \,\rm mm$     & GOODS-N            &  0.068 & 0.96-1.16 & 18   \\
\cite{weiss09} & LABOCA/APEX $870\,\rm\mu m$   & ECDF-S             &  0.25  & 1.2       & 19.2 \\            
\cite{aust10}  & AzTEC/JCMT  $1.1 \,\rm mm$    & LHE \& SXDF        &  0.5   & 1         & 18   \\
\cite{scot10}  & AzTEC/ASTE  $1.1 \,\rm mm$    & GOODS-S            &  0.14  & 0.48-0.73 & 30   \\  
\cite{hats10}  & AzTEC/ASTE  $1.1 \,\rm mm$    & AKARI, SXDF, \& SSA & 0 .25  & 0.32-0.71 & 30    \\
\hline
\textit{This Work} & MAMBO $1.2\,\rm mm$       & LHN                &  0.16  & 0.75      &   11
\enddata
 \label{t-areas}
\end{deluxetable}

\begin{deluxetable}{llccl}
\tablecaption{20\,cm counterpart identification rates}
\tablewidth{0pt}
\tablehead{ \colhead{} & \colhead{} & \colhead{} & \colhead{$\sigma_{1.4\,\rm GHz}$} & \colhead{} \\
            \colhead{Reference}          & \colhead{Instrument} &           \colhead{Field}      & \colhead{($\mu\rm Jy\,beam^{-1}$)}        & \colhead{ID rate}
}
\startdata
\cite{borys04} & SCUBA $850\,\rm \mu m$  & GOODS-N      & 9.0       & $58\%$ (11/19)    \\ 
\cite{ivis07}  & SCUBA $850\,\rm \mu m$  & LHE \& SXDF  & 4.2 \& 7 & $52\%$ (62/120)    \\
\cite{bert07}  & MAMBO $1.2\,\rm mm$     & COSMOS       & 8.5      & $73\%$ (11/15)     \\
\cite{scot08}  & AzTEC $1.1 \,\rm mm$    & COSMOS       & 10.5     & $44\%$ (12/47)     \\
\cite{chapi09} & AzTEC $1.1 \,\rm mm$    & GOODS-N      & 4.5      & $76\%$ (22/29)     \\
\cite{bigg10}  & LABOCA $870\,\rm\mu m$  & ECDF-S       & 6.5      & $37\%$ (47/126)   \\
\hline
\textit{This Work} & MAMBO $1.2\,\rm mm$ & LHN          & 2.7      & $93\%$ (38/41)  

\enddata
 \label{t-counterparts}
\end{deluxetable}

\clearpage

\clearpage

\end{document}